# A Data-Driven Travel Mode Share Estimation Framework based on Mobile Device Location Data

Mofeng Yang[1], Yixuan Pan[2], Aref Darzi[3], Sepehr Ghader[4], Chenfeng Xiong[5] and Lei Zhang[6,*]

1. Graduate Research Assistant (mofeng@umd.edu), ORCID: 0000-0002-0525-7978
2. Graduate Research Assistant (ypan1003@umd.edu), ORCID: 0000-0002-5015-0088
3. Graduate Research Assistant (adarzi@umd.edu), ORCID: 0000-0003-2558-5570
4. Research Scientist (sghader@umd.edu), ORCID: 0000-0003-1938-7914
5. Assistant Research Professor (cxiong@umd.edu), ORCID: 0000-0003-4237-1750
6. Professor (lei@umd.edu), * Corresponding Author, ORCID: 0000-0002-3372-6321

1-6: Maryland Transportation Institute (MTI), Department of Civil and Environmental Engineering, 1173 Glenn Martin Hall, University of Maryland, College Park MD 20742, USA.
5: Shock Trauma and Anesthesiology Research (STAR) Center, School of Medicine, 685 W Baltimore Street, Suite 600, University of Maryland, Baltimore MD 21201, USA.

## Abstract

Mobile device location data (MDLD) contains abundant travel behavior information to support travel demand analysis. Compared to traditional travel surveys, MDLD has larger spatiotemporal coverage of the population and its mobility. However, ground truth information such as trip origins and destinations, travel modes, and trip purposes are not included by default. Such important attributes must be imputed to maximize the usefulness of the data. This paper targets at studying the capability of MDLD on estimating travel mode share at aggregated levels. A data-driven framework is proposed to extract travel behavior information from MDLD. The proposed framework first identifies trip ends with a modified Spatiotemporal Density-based Spatial Clustering of Applications with Noise (ST-DBSCAN) algorithm. Then three types of features are extracted for each trip to impute travel modes using machine learning models. A labeled MDLD dataset with ground truth information is used to train the proposed models, resulting in 95% accuracy in identifying trip ends and 93% accuracy in imputing the five travel modes (drive, rail, bus, bike and walk) with a Random Forest (RF) classifier. The proposed framework is then applied to two large-scale MDLD datasets, covering the Baltimore-Washington metropolitan area and the United States, respectively. The estimated trip distance, trip time, trip rate distribution, and travel mode share are compared against travel surveys at different geographies. The results suggest that the proposed framework can be readily applied in different states and metropolitan regions with low cost in order to study multimodal travel demand, understand mobility trends, and support decision making.

**Keywords**: Travel mode share; travel surveys; machine learning; mobile device location data.

# 1. Introduction

Accurate measurement of travel behavior can help agencies understand how travel demand evolves and better allocate resources in support of transportation planning processes. Traditionally, researchers and practitioners design and conduct travel surveys to obtain household- and individual-level travel behavior information, including trip origins and destinations, trip distance, trip time, trip purposes, travel modes, etc. Two of the most widely used travel surveys conducted in the United States (U.S.) are the National Household Travel Survey (NHTS, see U.S. Department of Transportation 2017) and American Travel Survey (Lapham 1995). Methods to conduct travel surveys usually require respondents to record their daily trips with original paper-and-pencil interview (PAPI), computer-assisted telephone interview (CATI), and computer-assisted-self-interview (CASI) (Wolf et al 2001; Wolf 2006). However, these methods are prone to several well-known biases, such as under-reported short trips, inaccurate travel times, and travel distances (Stopher et al. 2007; McGowen and McNally 2007). Also, traditional travel surveys require complex planning and design, and large human labor and costs, and can only obtain relatively small survey samples for a limited number of cross-sections. For instance, if half of the 350 Metropolitan Planning Organizations (MPO) in the U.S. conduct travel surveys only once in a decade, it will result in $ 7.4 million per year cost (Zhang and Viswanathan 2013).

In the past two decades, along with the technological advancement in mobile sensors and mobile networks, mobile device location data (MDLD) has been growing dramatically in terms of data coverage and data size. In the realm of transportation, the abundant individual movement information stored in MDLD has great potential to help researchers and practitioners understand the bigger picture of human travel. Compared to traditional travel surveys, MDLD has larger spatial, temporal, and population coverage and it comes from various sources, including Global Positioning Service (GPS) devices, cellular network, Bluetooth, Wi-Fi, etc. To fully take advantage of MDLD, appropriate steps and methods need to be developed to extract useful travel behavior information from MDLD (Schönfelder et al. 2002; Axhausen et al. 2003).

This study aims to develop a data-driven framework to estimate travel mode share based on MDLD. The proposed framework is trained using a labeled MDLD dataset collected from a mobile application and is further applied to two large-scale MDLD datasets. The estimated trip distance, trip time, trip rate distribution, and travel mode share are compared with travel surveys. Results suggest that the proposed framework can be readily applied in many regions with low cost to obtain travel mode share estimates and study travel trends, which can help decision-makers prioritize multimodal travel needs. The remainder of the paper is organized as follows. Section 2 reviews the literature on MDLD. Section 3 describes the proposed data-driven framework and models in detail. Section 4 introduces the data. Section 5 presents the model development results. Section 6 demonstrates the framework with two large-scale case studies. Section 7 concludes this paper and discusses the limitations and future research directions.

## 2. Literature Review

In this literature review, we first introduce three types of MDLDs: GPS data, cellular data, and Location-based Service (LBS) data and review the state-of-the-practice applications for each of them. Then, we review the state-of-the-art methods for extracting trips and imputing travel modes from MDLDs.

### 2.1. Mobile Device Location Data

### 2.1.1. GPS Data

The GPS data in this study mainly refers to the personal longitudinal location data that are collected to enhance the quality of travel surveys with user recalled information. In the late 1990s, the GPS data logger was installed in the vehicle and charged by the vehicle battery (Battelle 1997; NuStats 2002; NuStats 2004; Ojah and Pearson 2008; Wolf and Lee 2008; ETC Institute 2009; ETC Institute 2011; ETC Institute 2011; ETC Institute 2011; ETC Institute 2011). The vehicle location was recorded seconds by seconds when the vehicle is moving (Ojah and Pearson 2006). This approach was proved to be effective, but it only captured vehicle trips. Later, the wearable GPS further allowed respondents to carry them such that trips traveled by other non-vehicle travel modes could also be recorded (Twin Cities Metropolitan Council 2012; Delaware Valley Regional Planning Commission 2013; Westat 2014; Westat 2015). Some travel surveys utilized both in-vehicle and wearable GPS data loggers to take advantage of both technologies (NuStats 2007; NuStats 2011; NuStats 2013). There's also another type of GPS data that is collected without any user recalled information using in-vehicle GPS devices for both passenger vehicles and trucks. For instance, INRIX Traffic collects GPS probe data from commercial vehicle fleets, connected vehicles, and mobile device applications (INRIX 2020). The data can be further aggregated into link- or corridor- level to provide a real-time estimation of traffic speed and travel time (Ali et al. 2009; Schrank et al. 2015; Cui et al. 2018).

### 2.1.2. Cellular Data

There is also another type of MDLD called cellular data, including Call Detail Record (CDR) and sightings (Chen et al. 2016). Call Detail Record (CDR) data is generated when a phone communicates with the cell tower in the cellular network, for instance when a phone call or a text message is made by the phone. The location information of CDR data is the cell tower locations thus it fully depends on the density of the cellular network and does not reflect the actual location of the device (Chen et al. 2016). Similar to CDR data, sightings are also generated through communication with cell towers but the actual location of the device is calculated via triangular calculation (Chen et al. 2016). Both types of cellular data have been widely used in studying human mobility patterns in the past two decades (Gonzalez et al. 2008; Kang et al. 2012; Kang et al. 2012).

### 2.1.3. Location-based Service Data

The Location-based Service (LBS) data is generated when a mobile application updates the device's location with the most accurate sources, based on the existing location sensors such as Wi-Fi, Bluetooth, cellular tower, and GPS (Chen et al. 2016; Wang and Chen 2019). The LBS data can reflect the exact location of mobile devices and thus provide invaluable location information describing individual-level mobility patterns. In most cases, the LBS data has a higher spatial precision and smaller Location Recording Interval (LRI) than the CDR data (Chen et al. 2016; Gonzalez et al. 2008; Kang et al. 2012; Kang et al. 2012; Wang and Chen 2019; Wang et al. 2019). Lots of applications have been developed using the LBS data. For instance, a recent smartphone-enhanced travel survey conducted in the U.S. used a mobile application, rMove developed by Resource Systems Group (RSG), to collect high-frequency location data and let respondents recall their trips by showing the trajectories in rMove (RSG 2014; RSG 2015; RSG 2015; RSG 2017); Airsage leveraged LBS data to develop a traffic platform that can estimate traffic flow, speed, congestion and road user sociodemographic for every road and time of day (Airsage 2020); Maryland Transportation Institute (MTI) at the University of Maryland (UMD) developed the COVID-19 Impact Analysis Platform (*data.covid.umd.edu*) to provide insight on COVID-19's impact on mobility, health, economy and society across the U.S. (Zhang et al. 2020)

In summary, these three types of MDLD are different in terms of spatiotemporal coverage of population and its mobility, and data quality, e.g., spatial accuracy and LRI. The GPS data usually has the highest spatial accuracy (e.g., 10 meters) and the lowest LRI (usually 1 second), but it usually covers a small percentage of the population, and thus cannot reflect population-level travel behavior without a statistical weighting process. The cellular data and LBS data have significantly higher spatiotemporal coverage of population over the GPS data. However, the ground truth information is usually missing and the LRI for both types of data is based on mobile device usage on telecommunication or location-based services and thus has larger variation. In addition, the cellular data has much lower spatial accuracy (e.g., 1 kilometer) and the LBS produces location data with a larger range of spatial accuracy depending on the location sensor utilized to generate each record.

## 2.2. Extracting Trips from Mobile Device Location Data: State-of-the-Art Methodologies

The trip end identification algorithm for low-LRI MDLDs, i.e., GPS data, has been well-studied and used in practical applications. To obtain accurate trip ends, the traditional way is the rule-based trip end identification methods. This type of method designs rules and parameters based on domain knowledge. The trip ends are obtained by applying the rules to location data point by point and at the same time examining the intra-relationship between several consecutive location points. The parameters used in these rules are mostly defined by domain knowledge, such as dwell time, speed, etc. (McGowen and McNally 2007; Gong et al. 2014; Axhausen et al. 2003; Tsui et al. 2006;

Bothe and Maat 2009; Stopher et al. 2005; Du and Aultman-Hall 2007; Stopher et al. 2008; Schuessler and Axhausen 2009; Gong et al. 2012; Assemi et al. 2016; Patterson et al. 2016). In recent years, some researchers also leveraged the supervised machine learning models as a supplement to the rule-based methods, which classify each location point as static or moving (Gong et al. 2015; Zhou et al. 2016; Gong et al. 2018). Different clustering methods were also applied to obtain trip ends by first identifying people's activity locations from the location data (Zhou et al. 2007; Chen et al. 2014; Ye et al. 2009; Yao et al. 2019). A recent study utilized a spatiotemporal clustering method with three combined optimization models to detect trip ends (Yao et al. 2019). In recent years, there is also a special focus on deriving the trip ends from LBS data. A "Divide, Conquer and Integrate" (DCI) framework was proposed to process the LBS data to extract mobility patterns in the Puget Sound region (Wang et al. 2019). The proposed framework combined a rule-based method and incremental clustering method to handle the bi-modally distributed LBS data. The results were aggregated at census tract-level and compared with household travel surveys (Wang et al. 2019).

After the trip ends are identified, it is also important to impute the travel mode for each trip to obtain multimodal travel patterns. Travel mode imputation can be categorized into mainly two approaches: (1) trip-based approach; and (2) segment-based approach. The trip-based approach is based on the already identified trip ends, where each trip has only one travel mode to be imputed. The segment-based approach separates the trip into fixed-length segments (time or distance) and then imputes the travel mode for each segment. Then the segment with the same travel mode will be further merged to form a single-mode trip. This study mainly considers the trip-based approach because the purpose of this study is to estimate travel mode share aggregated from individual trips. Table 1 summarizes typical methods for travel mode imputation using the trip-based approach.

Table 1. Literature Review on Travel Mode Imputation Methods.

| Author | LRI | Model | Features | Modes | Acc. |
|--------|-----|-------|----------|-------|------|
| Gong et al. 2012 | / | Rules | Speed, Acceleration, Transit Stations, Transit Network | Drive, Train, Bus, Walk, Bike, Static | 82.6% |
| Stenneth et al. 2011 | 30 s | RF | Speed, Acceleration, Heading change, Bus location, Transit Network | Drive, Bus, Train, Walk, Bike, Static | 93.7% |
| Bruunauer et al. 2013 | 1-10 s | MLP | Speed, Acceleration, Bendiness | Drive, Bus, Train, Walk, Bike | 92.0% |
| Xiao et al. 2015ss | 1 s | BN | Speed, Acceleration, Trip Distance | Drive Bus, Walk, Bike, E-Bike | 92.0% |

*\* RF: Random Forest; MLP: Multi-Layer Perceptron; BN: Bayesian Network.*

It can be observed that some typical features used are speed and acceleration (Gong et al. 2012; Stenneth et al. 2011; Brunauer et al. 2013; Xiao et al. 2015; Broach et al. 2019; Shafique and Hato 2016; Wang et al. 2017). Specifically, when the LRI is less than 10 seconds, the speed and acceleration features are more important to differentiate between different travel modes, which can be imputed solely by the data itself. When the LRI is relatively high, for instance, 30 s, additional features can be added to maintain the same level of accuracy such as real-time transit information (Stenneth et al. 2011), multimodal transportation network (Gong et al. 2012; Stenneth et al. 2011), sociodemographic information (Wang et al. 2017) etc.

Both the state-of-the-practice applications and the state-of-the-art methodologies are able to accurately identify trip ends and impute travel modes based on low-LRI GPS data with user-labeled ground truth information. However, most previous studies both trained and tested the algorithms on a small GPS data sample without evaluating their performances in large-scale and real-world MDLD datasets. In addition, limited efforts focused on developing suitable travel mode imputation algorithms for LBS data and validating the results. This study aims to fill this gap by proposing a data-driven travel mode share estimation framework based on LBS data and further implement it on large-scale datasets. The proposed framework investigates the spatial accuracy and LRI distributions of the LBS data sample and fine-tuned the algorithms considering their variations. The validation efforts based on the external data sources, such as household travel surveys, have proven the proposed framework efficient and versatile for applications.

## 3. The Data-Driven Travel Mode Share Estimation Framework

Figure 1 shows the proposed data-driven framework. On the left is the Model Development pillar, wherein a dedicated ground-truth data collection of labeled and mode-specific trips and trajectories is conducted in order to train the trip end identification algorithm and travel mode imputation model. These trained models are then applied to the Model Application pillar on the right. The Model Application generates trip rosters with imputed travel modes for the unlabeled MDLD datasets in the application contexts. Finally, a validation process compares the aggregated mode share, as well as other statistics, with travel surveys before the data products are deemed useful and applicable for any transportation planning applications.

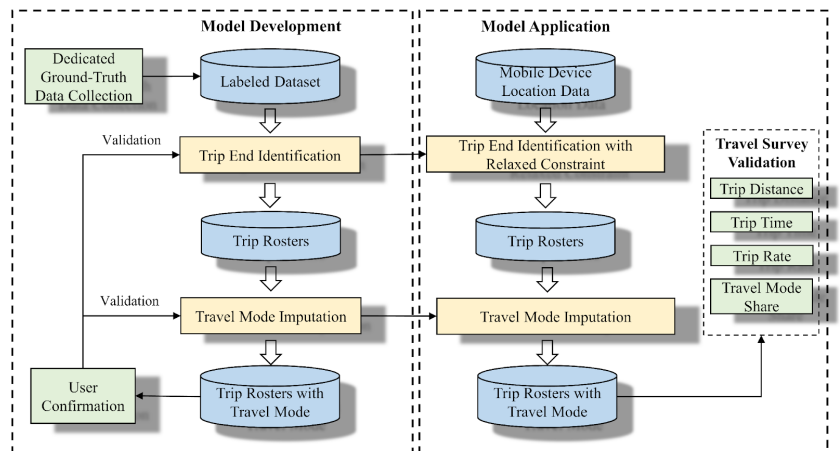

Figure 1. The Data-Driven Travel Mode Share Estimation Framework

### 3.1. Trip End Identification

Considering a person's daily travel, it is very common that he or she makes multiple stops for different trip purposes. As illustrated in Figure 2, these stops are categorized into two categories, namely Activity Stops (AS) and Non-Activity Stops (NAS). ASs represent stops where actual activities take place, such as home, workplace, restaurant, shopping mall, etc. NASs represent stops where no activity takes place or the activity takes a very short amount of time, usually including stopping at a traffic light, picking up people within a short range of time, etc. In this study, only ASs are considered as actual trip ends and the trajectory between two consecutive ASs is considered as a valid trip.

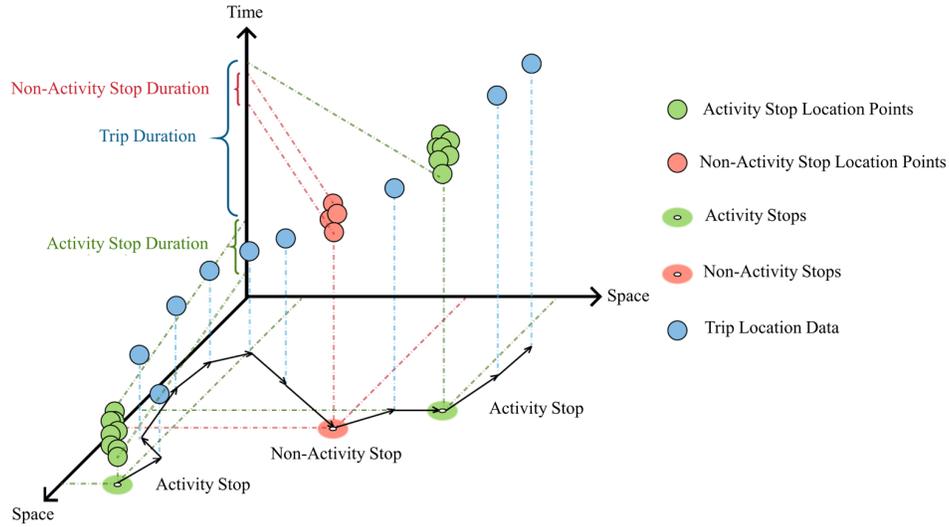

Figure 2. Typical Daily Travel Pattern of an Individual

The first step is to identify all stop points including all ASs and NASs. The Spatiotemporal Density-based Spatial Clustering of Applications with Noise (ST-DBSCAN) (Birant and Kut 2007) is applied to fulfill this step. The ST-DBSCAN is an extended version of the traditional DBSCAN algorithm (Ester et al. 1996) with consideration of both spatial and temporal constraints. The temporal constraint was able to handle the scenarios when a person visits the same location multiple times per day, i.e., home, work. Three thresholds are defined for the ST-DBSCAN used in this study: (1) the spatial threshold $s$: it represents the distance falling within the activity distance range; (2) the temporal threshold $t$: it represents the minimum duration of an activity; and (3): the minimum neighbor's $m$: it represents the minimum number of location points to form a cluster. Details of ST-DBSCAN can be found in Birant and Kut (2007) and Ester et al. (1996).

With all stop points identified, the second step is to distinguish between ASs and NASs. Two parameters are proposed: (1) $s_{act}$: maximum activity distance range. If the distance between two consecutive clusters stayed within $s_{act}$, it implies that these two clusters might still belong to the same activity, and the location points falling within these two clusters would be labeled as activities, otherwise a trip will be generated. (2) $t_{act}$: minimum activity duration threshold of an activity. If the minimum time lag between the last stop point and the first stop point of two consecutive clusters is shorter than $t_{act}$, it implies no activity happens, which can happen at traffic lights, traffic congestions, pick up, etc, otherwise an activity would be identified between the two clusters.

### 3.2. Travel Mode Imputation with Machine Learning Models

This study proposes machine learning models to impute five travel modes (drive, bus, rail, bike and walk) and four travel modes (drive, bus, rail, and non-motorized) from trips identified from the previous step. Five machine learning models will be examined in terms of prediction accuracy, including K-Nearest Neighbors (KNN), Support Vector Classifier (SVC), eXtreme Gradient Boosting (XGB), Random Forest (RF), and Deep Neural Network (DNN). The five machine learning modes are implemented using the scikit-learn package in Python (Pedregosa et al. 2011). A detailed introduction of these models can be found in Appendix I.

Table 2. Features Constructed from Mobile Device Location Data

| Features | Unit |
| --- | --- |
| *Location Recording Interval Feature* | |
|     Average # of records per minutes | frequency |
| *Trip Features* | |
|     Trip distance | meters |
|     Origin-Destination distance | meters |
|     Trip time | minutes |
|     Max., Min., Avg, Med., 5-, 25-, 75-, 95- percentile speed | meter / second |
| *Multimodal Transportation Network Features (location data with accuracy < 50 meters)* | |
|     % of location points fell within 50 meters of rail network | percentage |
|     % of location points fell within 50 meters of bus network | percentage |
|     % of location points fell within 50 meters of drive network | percentage |
|     % of location points fell within 50 meters of bus stops | percentage |

Feature set construction directly affects the model performances. Three types of features, LRI feature, trip features, and multimodal transportation network features, are constructed from MDLD, as shown in Table 2. The LRI feature, represented by the average number of records per minute, indicates the location service usage during a trip. The trip features can show the characteristics of each trip, including trip distance, origin-destination distance, trip time, average speed, minimum speed, maximum speed, median speed, and 5, 25, 75, 95- percentile speed. The multimodal transportation network features are important to distinguish between different travel modes (Bohte and Maat 2009; Gong et al. 2018). Here, a 50-meter buffer for the multimodal transportation networks (rail, bus, and drive), and bus stops are generated to obtain the percentage of location points for each trip that fall within each buffer respectively. It should be noted that since the accuracy of these multimodal network features is largely affected by the spatial accuracy of the location data, we used only location data with spatial accuracy less than 50 meters to calculate these features to obtain more accurate estimates. Accelerator meter data is also useful in travel mode imputation. However, it was not taken into account because such information is generally not available in large-scale MDLD datasets purchased from the third-party vendors.

## 4. Data

### 4.1. Mobile Device Location Data

This study uses three MDLD datasets that can be categorized into LBS data. The first LBS dataset is collected from a mobile application called incenTrip (*incentrip.org*). incenTrip was developed by Maryland Transportation Institute (MTI) at the University of Maryland (UMD) to nudge travel behavior changes by providing real-time dynamic incentives in the Washington Metropolitan Area (Xiong et al. 2019). incenTrip collects location data using the Google Maps API (Android) or Apple Core Location (iOS) with a pre-defined fixed LRI and inserts the data into PostgreSQL in the Amazon Web Services (AWS) to ensure privacy protection. The LRI of the incenTrip data is set to be 1-15 seconds with three considerations: 1) help preserve battery of the mobile devices, and at the same time fully capture users daily travel behavior; 2) ensure the capability of data for travel mode imputation, as the literature suggests, reducing the LRI of the data will decrease the travel mode imputation accuracy (Shafique and Hato 2016); and 3) improve similarity with large-scale LBS datasets. Each record of the raw location data includes a unique device identifier, latitude, longitude, and timestamp information. The proposed framework would then be applied to identify trips, impute the corresponding travel modes, and then update the trip records with imputed modes in the database. This study uses the incenTrip app from March 2019 to January 2020 for a dedicated ground-truth MDLD data collection. During the 10-month period, fifteen designated respondents of all the incenTrip users were hired to travel with incenTrip and record detailed information for each trip daily, including the start date, start time, end date, end time, origin street address, destination street address, travel time and travel mode. The trip identification algorithm is calibrated by comparing their travel diaries and the algorithm outputs. In addition, we also collected the trips with confirmed travel modes by other users. As a result of this data collection effort, a total number of 12,688 ground-truth trip records with travel mode labels were obtained from 410 users for the subsequent travel mode imputation model training process. Figure 3 visualizes the trajectories of these trips in different colors by travel modes.

Two other LBS datasets are obtained from one of the leading data vendors in the U.S. The data is generated by multiple mobile applications. In this study, the two LBS datasets are extracted with different spatial, temporal, and population coverages to keep a comparable data size. Table 3 describes these two datasets. Dataset I covers the Baltimore-Washington metropolitan area, including the state of Maryland, District of Columbia (D.C.), and Northern Virginia. It has a temporal coverage of one typical weekday, Sept. 12nd in 2017. A total of 474,634 unique devices observed in this area are considered. Dataset II expands the spatial coverage to the U.S. It has a temporal coverage of seven days from Aug. 1ˢᵗ to Aug. 7ᵗʰ in 2017. 3% of the total number of devices observed is randomly sampled and used for this study in order to keep a comparable data size, including 266,149 unique devices.

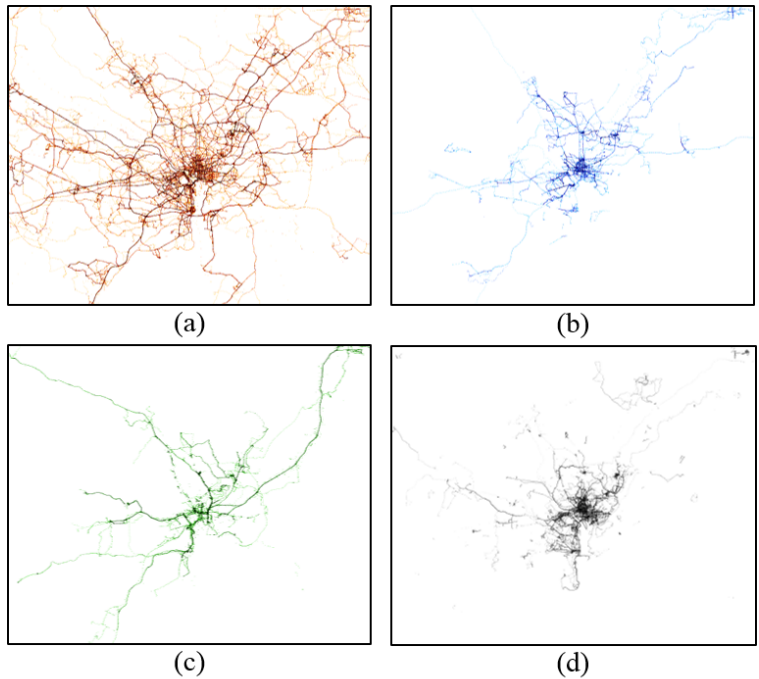

(a)  (b)

(c)  (d)

Figure 3. incenTrip Trip Trajectories for (a) drive; (b) bus; (c) rail; (d) non-motorized.

Table 3. Location-Based Service Data Description

| Dataset | Spatial Coverage | Temporal Coverage | Sample Rate | Sampled Numbers |
|---------|------------------|-------------------|-------------|-----------------|
| I | Baltimore-Washington metropolitan area | Sept. 12ⁿᵈ, 2017 | 100% | 474,634 |
| II | the United States | Aug. 1ˢᵗ – Aug. 7ᵗʰ, 2017 | 3% | 266,149 |

## 4.2. Multimodal Transportation Networks

This study also collects the multimodal transportation network data including drive, bus, rail networks, and bus stop locations to construct network-related features. The drive network is collected from the Highway Performance Monitoring System (HPMS) (FHWA 2020) that includes national freeway and arterial roads in the U.S. The national bus and rail network and the bus stops data are collected from the United States Department of Transportation (U.S. DOT) Bureau of Transportation Statistics (BTS) National Transit Map (NTM) (U.S. DOT BTS 2020). Figure 4 illustrates the multimodal transportation networks used in this study.

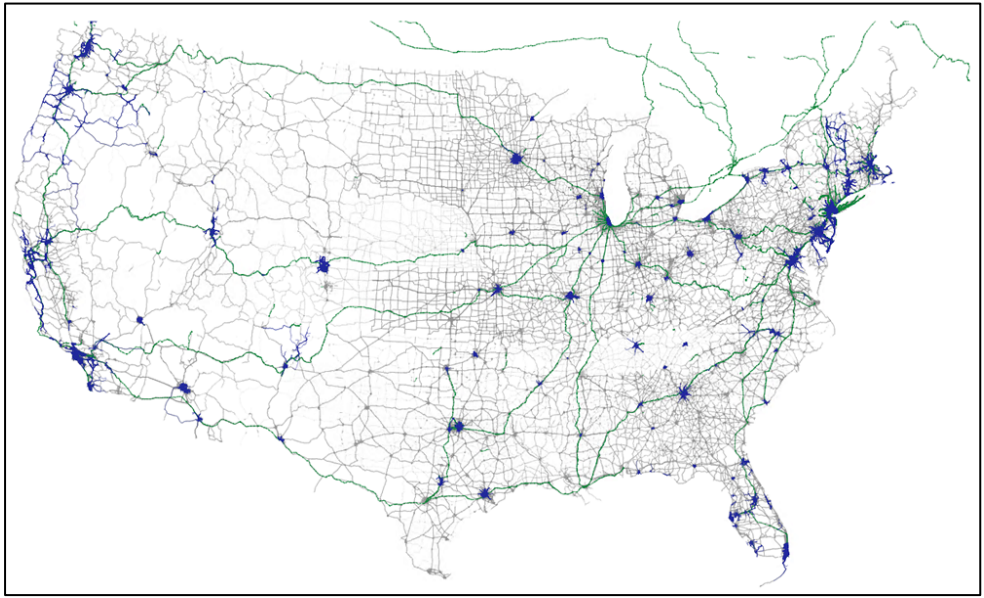

Figure 4. Multimodal Transportation Networks: drive (grey), rail (green), bus (blue).

### 4.3. Travel Surveys

Two travel surveys are used in this study for comparison purposes: NHTS 2017 and 2007/08 TPB-BMC Household Travel Survey (HHTS). NHTS 2017 is a national-level travel survey conducted by USDOT Federal Highway Administration (FHWA), collecting travel behavior data from U.S. residents. The NHTS 2017 includes a total number of 129,696 households covering all 50 states and the District of Columbia, including trip origin and destinations, trip time, trip purposes, and travel modes (U.S. DOT 2017). The 2007/2008 TPB-BMC HHTS is conducted by Transportation Planning Board (TPB) and Baltimore Metropolitan Council (BMC) in Baltimore and Washington regions from February 2007 to March 2008 using the same survey designs (MWCOG 2010). This survey covered nearly 14,000 households and can provide travel mode shares at Traffic Analysis Zone (TAZ) level. In our case study, we aggregated the travel mode shares at the county level using the trip origins in the travel survey and further compare with the LBS estimates.

### 5. Model Development

#### 5.1. Trip End Identification Result

Five parameters of the proposed ST-DBSCAN are calibrated using the incenTrip data: the spatial threshold $s$, temporal threshold $t$, minimum neighbors $n$, maximum distance threshold for an activity $s_{act}$, and minimum duration of an activity $t_{act}$. Since $s$ determines the distance range of a stop, increasing the value of $s$ would identify more stop points since more location points would be clustered. To ensure all the stop points are captured including traffic congestions and waiting at a traffic light for both vehicle and pedestrian, four constraints are added as shown below:

$$t \geq n \cdot f$$

$$t_{act} \geq n \cdot f$$

$$s_{act} \geq s$$

$$n \cdot f \geq s/v$$

where $v$ is the average walking speed, here we consider 1 m/s; $f$ is location recording interval. Consider the real-world scenario when a person stops, it is intuitive to set the $s$ value to be relatively small. Here we use 25-meter, 50-meter, and 100-meter as the candidate $s$ value. Also, the $t_{act}$ was set as 300 seconds to obtain most of the short activities. Then, with the given LRI the corresponding range for other parameters could be calculated. Table 4 shows the calibrated parameters used in the case study for each LRI.

Figure 5 shows the trip end identification result. The "Reported Trips" represents the trips that are reported in the respondents' travel diaries; the "Matched Trips" represents how many of the "Reported Trips" are identified from the data; the "All Identified Trips" represents all trips identified from the data, including the matched trips and the underreported trips that are not recorded in respondents' travel diary; and the "Hit-Ratio" is the value of "Matched Trips" divided by "Reported Trips". It should be noted that for the 1-s LRI data, only 23 reported trips are collected due to the short testing period. Over 90% of reported trips can be identified for each LRI, with the overall Hit-Ratio equals to 94.5%. In addition, about 15% to 35% of the underreported trips are identified and confirmed by testers. Capturing these underreported trips help produce more detailed travel patterns of each respondent.

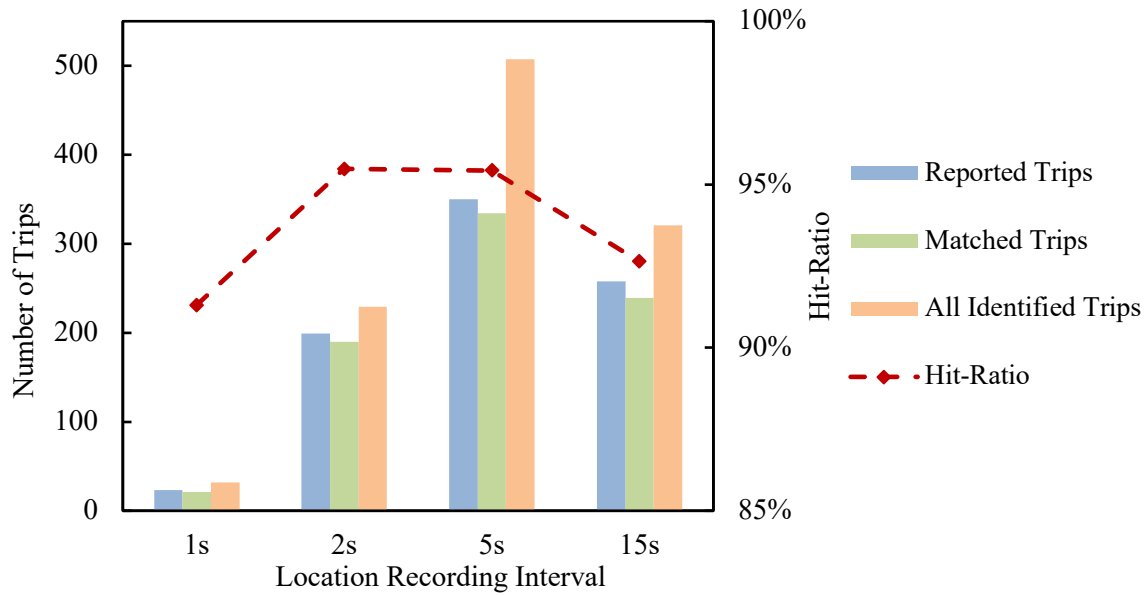

Figure 5. Trip End Identification Results.

Table 4. Calibrated Parameters for Each Location Recording Interval.

| LRI (s) | $s$ (m) | $t$ (s) | $n$ | $s_{act}$ (m) | $t_{act}$ (s) |
|---|---|---|---|---|---|
| 1 | 50 | 100 | 50 | 100 | 300 |
| 2 | 50 | 200 | 25 | 100 | 300 |
| 5 | 50 | 500 | 15 | 100 | 300 |
| 15 | 50 | 600 | 10 | 100 | 300 |

## 5.2. Travel Mode Imputation Result

Five machine learning models, KNN, SVC, XGB, RF, and DNN are used to impute travel modes. A total of 6,064 drive trips, 1,824 rail trips, 1,403 bus trips, 1,496 bike trips and 1,901 walk trips are collected from the incenTrip. 70% of the data is used for training and 30% of the data is used for testing. The Synthetic Minority Over-Sampling Technique (SMOTE) is then applied to the training data to address the imbalanced sample problem, where the minority class from the existing samples is synthesized (Chawla et al. 2002). For each machine learning method, the randomized search approach is used to fine-tune the model. Detailed hyperparameters can be found in Appendix I. During the model training process, 10-fold cross-validation (CV) is conducted to evaluate the model performance. The well-trained models are then applied to the testing data the F1 scores are calculated using the equations as shown below:

$$Precision = \frac{TP}{TP + FP}$$

$$Recall = \frac{TP}{TP + FN}$$

$$F1 = 2 \cdot \frac{Precision \cdot Recall}{Precision + Recall}$$

where TP represents the true positive, FP represents the false positive, and FN represents the false negative.

Table 5 compares the model performances using F1 scores using the testing data. It can be seen that RF achieved the highest CV-accuracy in both four and five modes models. The bus mode has the least prediction accuracy over the other modes. One possible reason might be the similarity between the drive trips and bus trips. As shown in Figure 3 (a) and (b), the drive trips cover most interstate highways, major arterials while the bus trips will only follow the pre-designed bus routes and will stop at the bus stops. However, when it comes to the roadway segments where bus routes exist, the moving patterns of passenger cars and buses highly depend on the traffic conditions and thus could share similar spatiotemporal characteristics, which requires high-quality data and complex method to distinguish with each other (Nguyen and Armoogum 2020). Therefore, in such roadways, especially in urban settings, the general level of LBS data quality might not be sufficient to capture the differences in their moving patterns, e.g., distinguishing between the short and frequent stops due to traffic congestion and bus stops. The RF model's feature importance can also be found in Appendix I.

Table 5. Model Performance Comparison.

| | | KNN | SVC | XGB | RF | DNN |
|---|---|---|---|---|---|---|
| Four Modes | Drive | 0.81 | 0.82 | 0.89 | 0.89 | 0.85 |
| | Rail | 0.87 | 0.91 | 0.91 | 0.91 | 0.89 |
| | Bus | 0.50 | 0.52 | 0.64 | 0.65 | 0.57 |
| | NonMotor | 0.76 | 0.85 | 0.89 | 0.89 | 0.86 |
| 10-Fold CV Accuracy | | 86.0% | 86.3% | 93.0% | 93.3% | 86.5% |
| Five Modes | Drive | 0.80 | 0.82 | 0.89 | 0.89 | 0.84 |
| | Rail | 0.87 | 0.90 | 0.91 | 0.91 | 0.90 |
| | Bus | 0.48 | 0.53 | 0.64 | 0.65 | 0.56 |
| | Bike | 0.50 | 0.69 | 0.77 | 0.79 | 0.73 |
| | Walk | 0.69 | 0.82 | 0.87 | 0.87 | 0.85 |
| 10-Fold CV Accuracy | | 85.5% | 85.4% | 93.0% | 93.6% | 85.0% |

## 6. Case Study

### 6.1. Comparison between the incenTrip Data and the Two LBS Datasets

Before we apply the calibrated framework to the two large-scale LBS datasets, the spatial accuracy and the LRI of these three datasets need to be compared. Figure 6 (a) shows the spatial accuracy of the three datasets after removing the data with an accuracy greater than 100 meters (see Appendix II. for more details). The incenTrip dataset has the best spatial accuracy among the three datasets, with more than 80% of the data's spatial accuracy less than 50 meters, whereas the two LBS datasets show a bimodal distribution, with the second peak locates around 70 meters. In general, the spatial accuracy of these three datasets is of high quality because, for all three datasets, around 90% of the data has a spatial accuracy of less than 100 meters. Figure 6 (b) only shows the LRI distribution for the two large-scale LBS datasets, since the incenTrip dataset is collected with pre-defined LRI, where a bimodal distribution can be observed (Wang et al. 2019). For each dataset, more than 75% of the data has LRI less than 15-s. Therefore, with consideration of both spatial accuracy and LRI, we applied the framework calibrated by the incenTrip data with 15-s LRI to estimate trips and travel mode shares for the two LBS datasets. In addition, since we observed that the second peak of the bimodal distribution locates around LRI = 120 s, two further relaxations are made in order to relax the restrictions of activity cluster identification: (1) the temporal threshold $t$ is relaxed from 600 s to 1800 s; (2) the minimum neighbors $n$ is relaxed from 10 to 5. Increasing the temporal threshold $t$ and decreasing the minimum neighbors $n$ can help capture a comparable number of activity clusters for LBS data with small LRI. Therefore, as shown in the previous calibration results (Table 4), data with high LRI requires larger $t$ and smaller $n$. the final parameters used for the two case studies are $s = 50$ m, $t = 1800$ s, $n = 5$, $s_{act} = 100$ m and $t_{act} = 300$ s. More analysis regarding the spatial accuracy and LRI of the LBS datasets used in this study can be found in Appendix II.

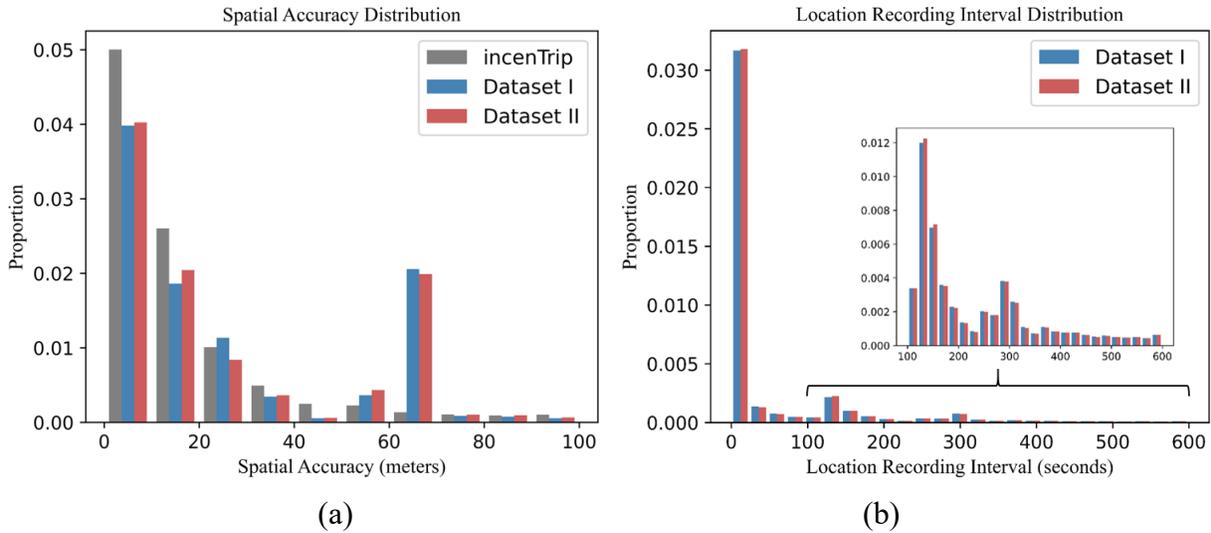

(a)                                              (b)

Figure 6. (a) Spatial accuracy distribution for the three LBS datasets; (b) Location recording interval distribution for the two case studies' LBS Datasets.

## 6.2. Case Study I: Baltimore-Washington Metropolitan Area Dataset

In the first case study, the proposed framework is applied to the LBS data observed in the Baltimore-Washington metropolitan area, covering the state of Maryland, D.C., and Northern Virginia. The trip distance, trip time and trip rate distribution are firstly compared with the 2007/08 TPB-BMC HHTS. The travel mode share is then compared at statewide- and county-level. A visualization of bus and rail travel distribution is provided at the census tract-level.

Trips are categorized into short-distance and long-distance trips using the 50 miles threshold (Hu and Reuscher 2004). Figure 7 (a) and (b) compare the short- and long-distance trip distance distribution with the 2007/08 TPB-BMC HHTS. For both short-distance and long-distance trips, a similar distribution can be observed, while more long-distance trips can be observed in the LBS data than the survey. Figure 7 (c) shows the trip time distribution comparison. The overall trend is similar, and still more long-duration trips can be observed in the LBS data. Figure 7 (d) compares the trip rate distribution, where the LBS data observed more devices with one trip per day. The number of trips by the time of day is also compared using average weekday trips from the survey, as shown in Figure 7 (e). It can be seen that both morning and afternoon peaks are matched in the two data sources, while LBS data can capture more daytime trips.

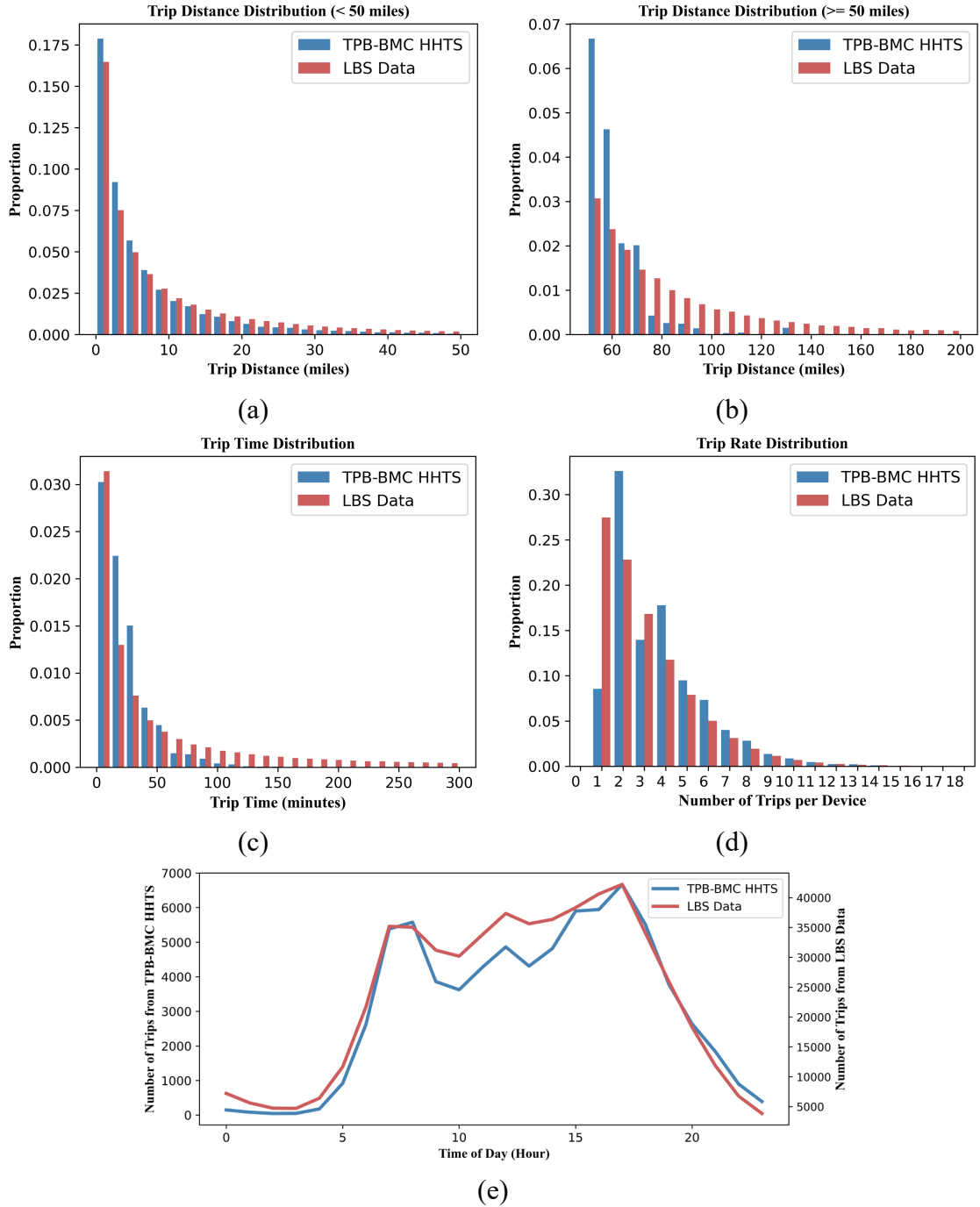

(a)

(b)

(c)

(d)

(e)

Figure 7. Comparison with 2007/08 TPB-BMC HHTS on: (a) Short-distance trips; (b) Long-distance trips; (c) Trip time; (d) Trip rate; (e) Number of trips.

Figure 8 (a) shows the statewide travel mode share comparison result. The overall mode share distribution is consistent with the 2007/08 TPB-BMC HHTS mode share, while the LBS data estimates lower bus mode share and higher non-motorized mode share. One reason might be people tend to underreport short-distance trips and most of the short-distance trips are walking. The travel mode share is further compared at the county-level, as shown in Figure 8 (b). The Pearson correlation between travel mode share estimated from the LBS data and the survey is calculated. Strong correlation can be observed, with the Pearson correlation value 0.98. Figure 8 (c) shows that for all eight counties in Maryland and D.C. It can be seen that the bus travel is underestimated, and the non-motorized mode share is overestimated from the LBS data. However, it should be noted that the 2007/08 TPB-BMC HHTS is conducted over ten years ago and the travel patterns might have been significantly changed. Therefore, we further compare the LBS data estimates with the recent NHTS 2017. However, due to the small sample size of NHTS 2017 in some rural counties (Kent County, Charles County, etc.), the NHTS 2017 estimates are prone to be over/underestimated. A detailed comparison can be found in Appendix II.

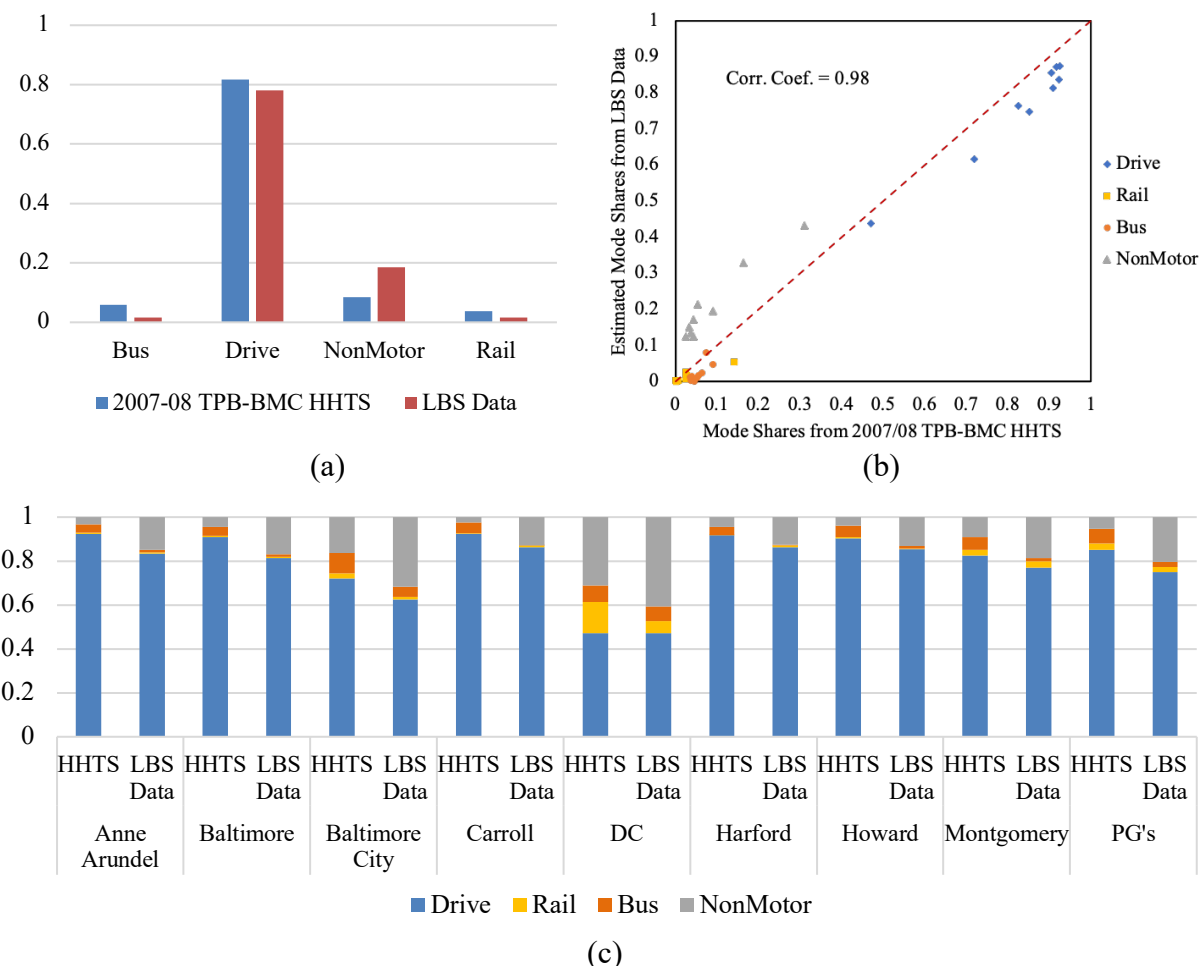

(a)

(b)

(c)

Figure 8. (a) Statewide travel mode share comparison; (b) County-level travel mode share correlation; (c) County-level travel mode share comparison.

Figure 9 plots the census tract-level rail and bus travel mode share using Jenks natural breaks optimization (Jenks 1967), with the deeper color representing higher mode share. For both D.C. and Baltimore city, the travel mode share distribution follows the geographical layout of rail and bus networks. Also, since D.C. has denser rail and bus networks, the relative mode share is higher than Baltimore City.

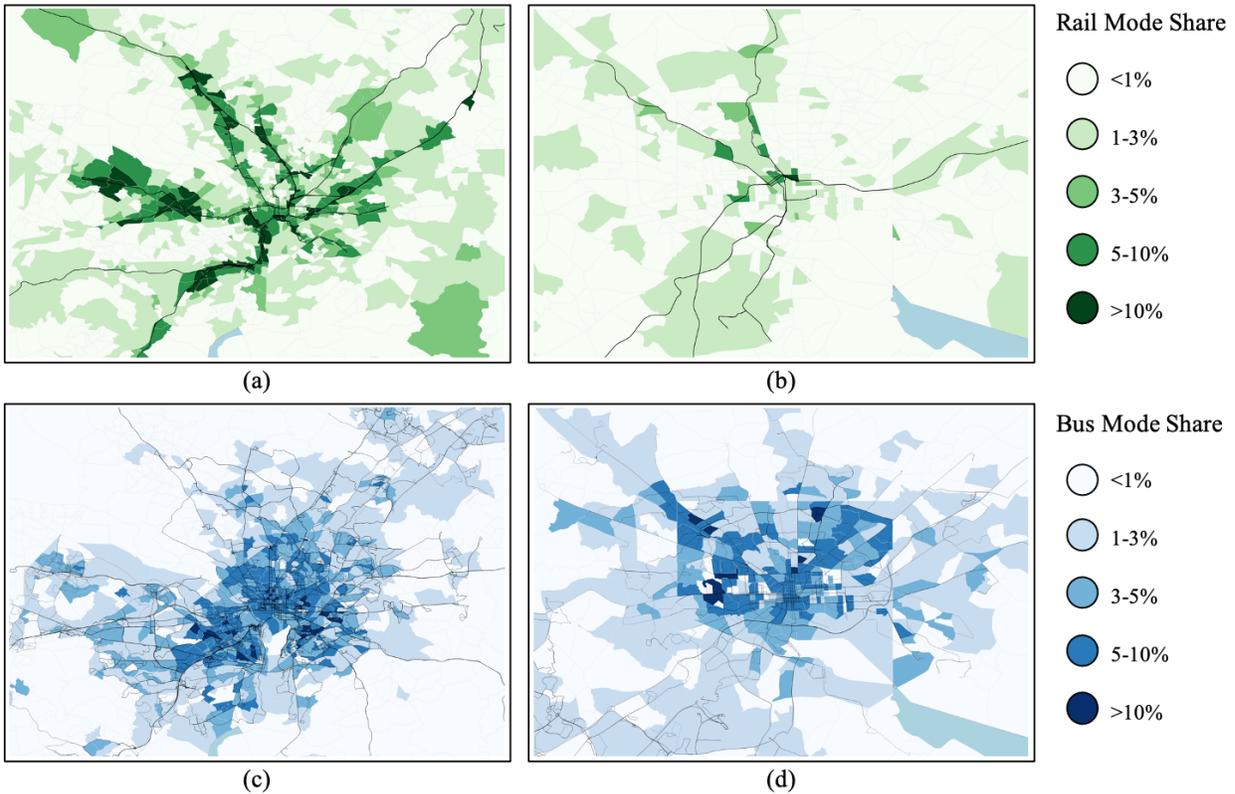

Figure 9. Census tract-level mode share illustration for (a) and (b): Rail mode share in D.C. and Baltimore City; (c) and (d): Bus mode share in D.C. and Baltimore City.

## 6.3. Case Study II: the U.S. National Dataset

In the second case study, the proposed framework is applied to the LBS data observed in the entire U.S. for a week, with 3% of the observed devices randomly sampled. The trip distance, trip time, trip rate distribution an,d travel mode share are compared against the NHTS 2017 at the nationwide- and state-level. A visualization is provided at CBSA-level.

Figure 10 (a) and (b) show the trip distance distribution comparison between NHTS results and LBS data results for short-distance and long-distance trips respectively. For both short-distance and long-distance trips, the overall trip distance distribution is consistent with the NHTS, while more long-distance trips can be observed. Figure 10 (c) shows the trip time distribution comparison. The overall trend is similar. The short trips are underestimated, and the long trips are overestimated. One possible reason for this observation is that NHTS 2017 calculated the network shortest path for trip distance, which is not always representative of the real path. Figure 10 (d) compares the daily trip rate to NHTS 2017, where the LBS data observed more devices with one trip per day. The daily variation within a week is also shown in Figure 11. The difference between weekends and weekdays is captured.

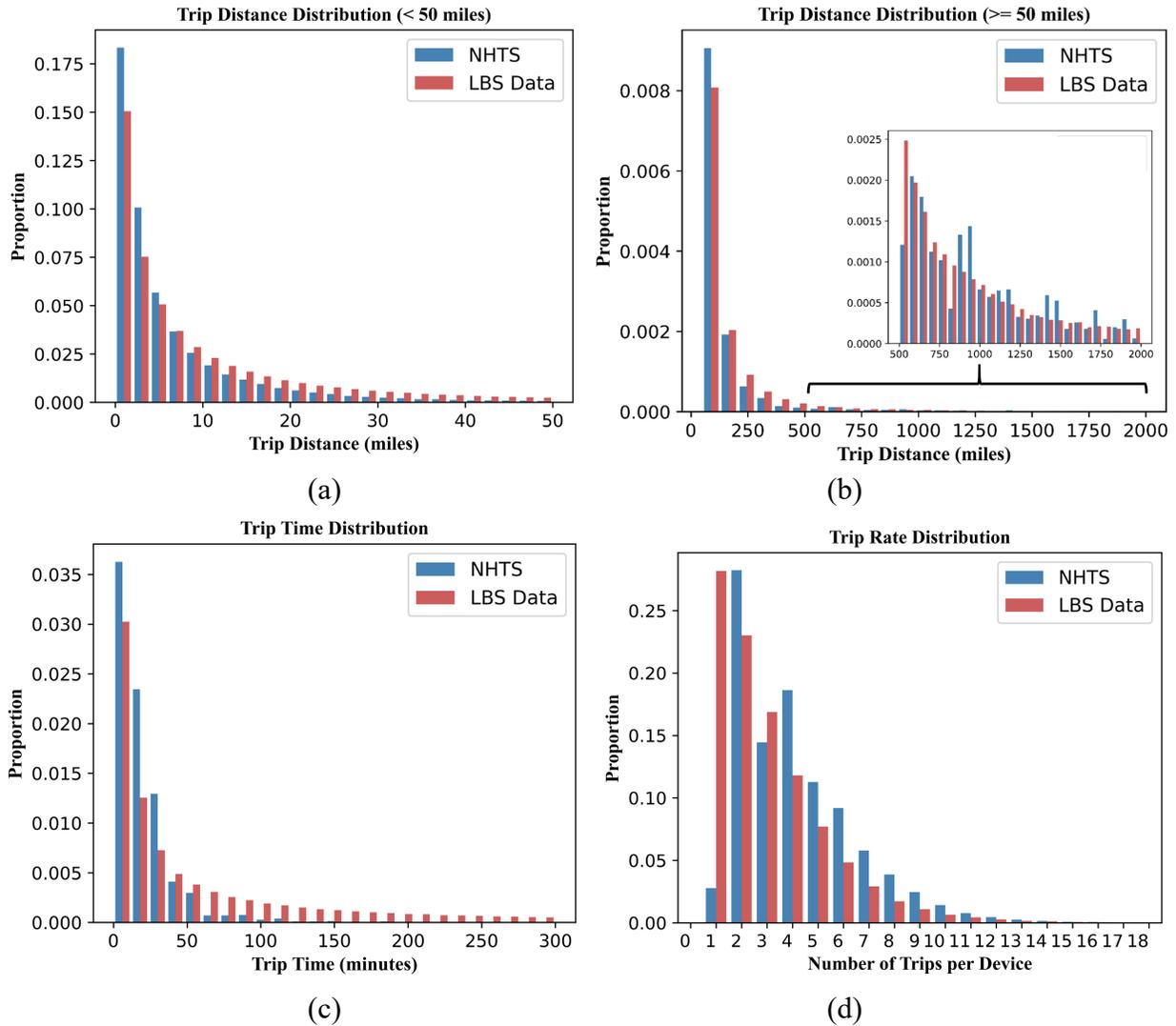

(a)

(b)

(c)

(d)

Figure 10. Comparison between NHTS and LBS data on: (a) Short-distance trips; (b) Long-distance trips; (c) Trip time; (d) Trip rate.

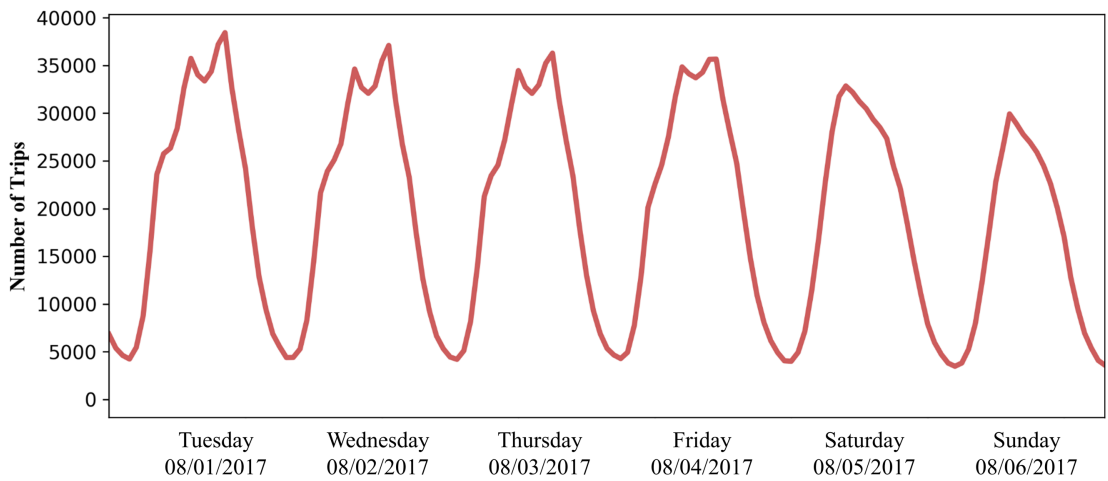

Figure 11. Daily Variation.

The air trips are firstly filtered out with a heuristic rule using three trip features: average speed, trip time, and trip distance. Here the 100 mph, 1 hour, and 100 miles are selected as the value of these thresholds, indicating that for a trip, if the average speed is larger than 100 mph, the trip time is larger than 1 hour and the trip distance is larger than 100 miles, then this trip is labeled as an air trip. Figure 12 shows the heat map of air trips' origins, where the depth of the color represents the number of air trips originated from the closest airport. It can be observed that almost all the major airports are captured.

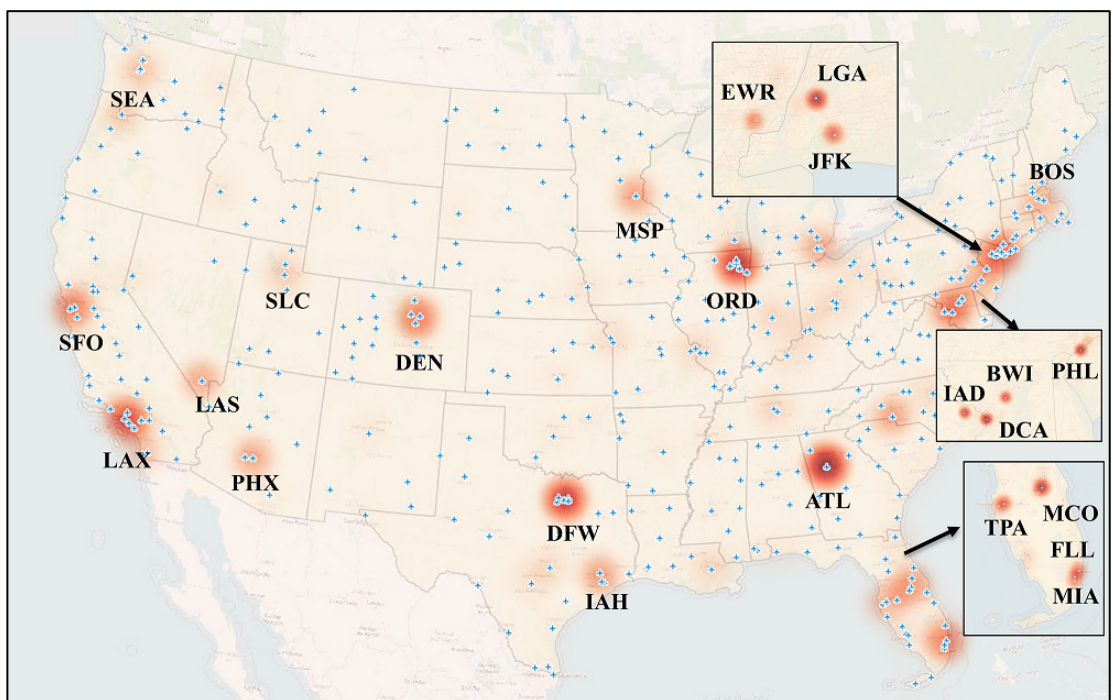

Figure 12. Nationwide-level Air Trip Origins

Figure 13 (a) shows the nationwide travel mode share comparison result. The overall mode share distribution is consistent with the NHTS 2017, with the bus mode share underestimated. Figure 13 (b) compares the travel mode shares at the state level across all 50 states and D.C. The Pearson correlation is also calculated between travel mode share estimated from the LBS data and NHTS 2017, where a strong correlation can be observed with the value of 0.99. To demonstrate the transferability of the proposed framework, eight states across the U.S. are selected for detailed comparison, as shown in Figure 13 (c). It can be seen that the overall travel mode share estimates are reasonable, with a slight underestimation of bus and rail travel and overestimation of the non-motorized travel. The results demonstrate the transferability and the generalization ability of the proposed framework that can be applied in larger geographies. Detailed comparison for all 50 states and D.C. can be found in Appendix II.

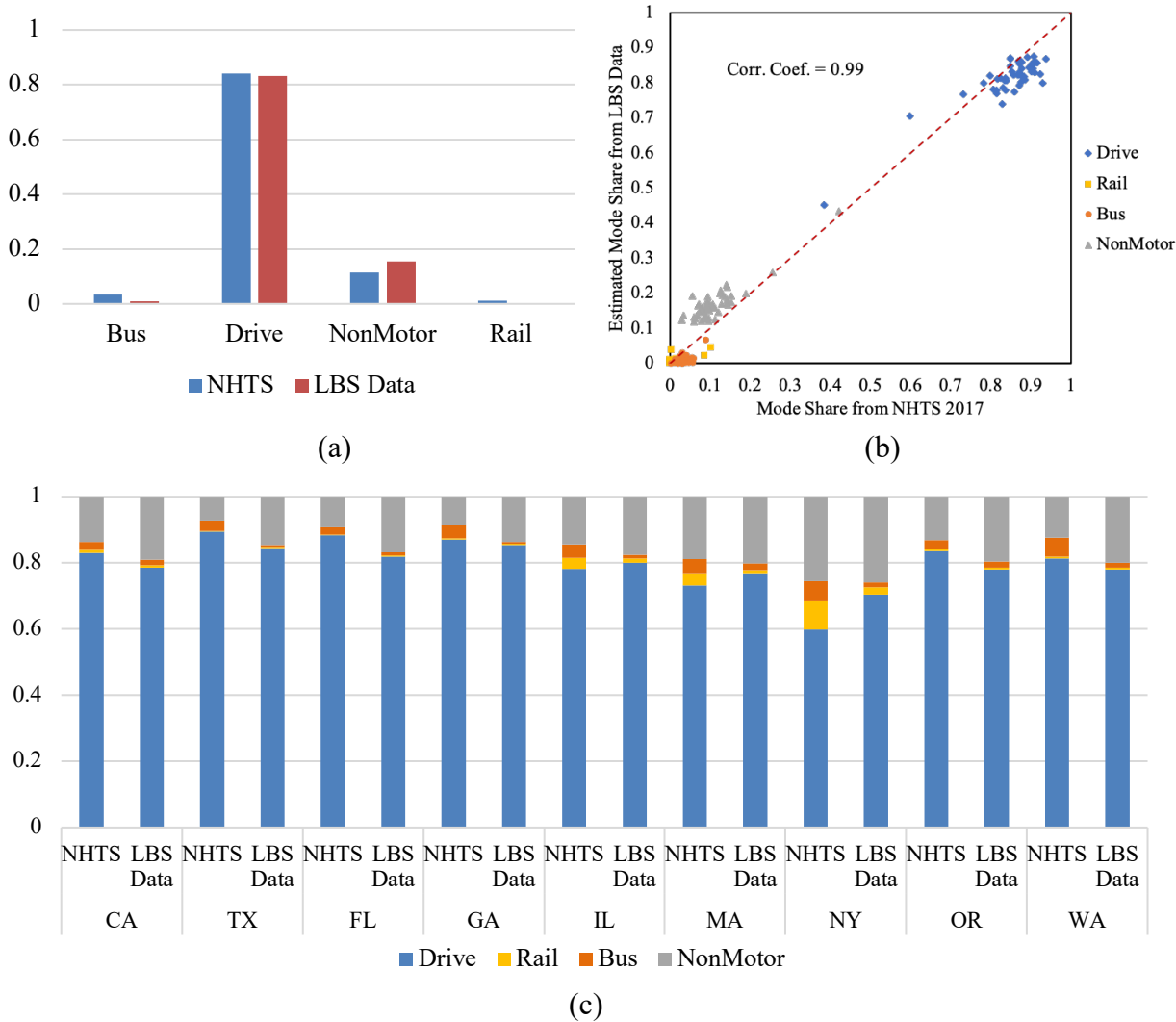

Figure 13. (a) Nationwide travel mode share comparison; (b) State-level travel mode share correlation; (c) State-level travel mode share comparison.

Figure 14 (a) and (b) visualize the CBSA-Level rail and bus travel mode share. The CBSAs with high bus and rail travel mode share estimates are those who have well-developed bus and rail networks. For instance, D.C., New York, Boston, and San Francisco have higher rail mode share than the other CBSAs. For bus mode share, a similar trend is observed too, where CBSAs with denser bus networks have higher bus mode share.

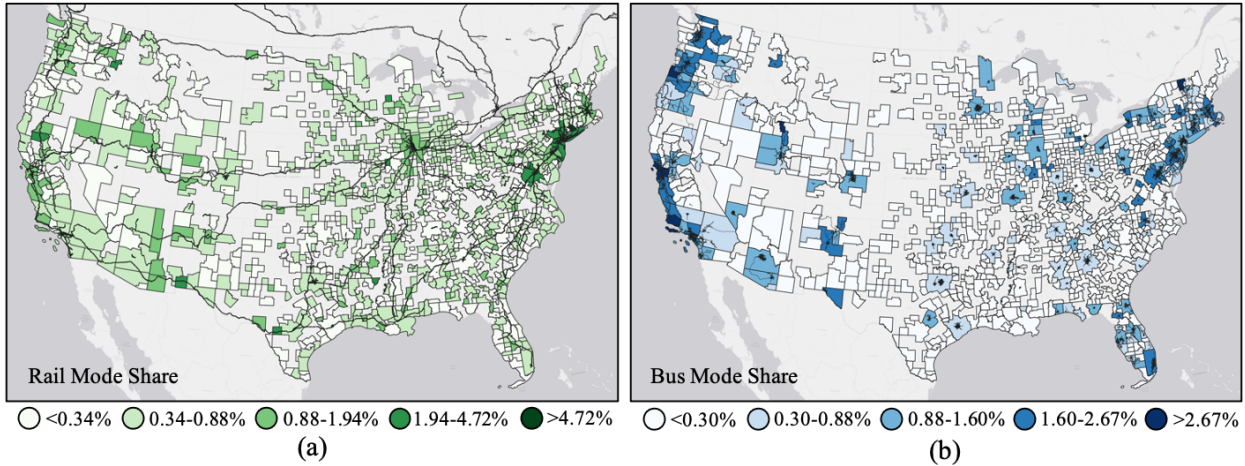

Figure 14. CBSA-level illustration of (a) rail travel mode share; (b) bus travel mode share.

# 7. Conclusions and Discussions

This study reviews the state-of-the-practice applications and state-of-the-art methods for extracting travel behavior information from MDLDs. Based on the literature review, the key research gap is identified, and a data-driven framework is proposed to estimate travel mode shares from MDLDs. The proposed framework reaches 95% accuracy in identifying trip ends and 93% accuracy in identifying five travel modes with the RF model. The developed framework is applied to two large-scale LBS datasets, covering the Baltimore-Washington metropolitan area and the U.S. with different spatial, temporal, and population coverage. The trip distance, trip time, and trip rate distribution, and travel mode share estimated from the two LBS datasets are compared with travel surveys for a comprehensive validation. The comparison results suggest that the proposed framework performs robustly in both geographical regions, indicating a good transferability.

One major challenge in the travel mode imputation is to distinguish different travel modes in urban settings, especially between drive and bus. This is also the reason why we solicited our training samples in the Washington Metropolitan Area. In the meantime, the multimodal networks are simpler in most rural areas: the rail network is sparse, the bus service is limited, etc. From our comparison with travel surveys at the county/state level, it shows that the current model trained with our incenTrip dataset could achieve satisfactory estimation in other contexts or regions.

The limitations of this study are summarized and discussed into three aspects: training data, validation data, and sample bias.

- **Training Data**: The proposed framework is calibrated and trained based on samples collected in the Washington Metropolitan Area with the incenTrip application. Even though the general travel mode share statistics are consistent with the travel surveys, in the real world, the travel behavior might be different from region to region, resulting in biased travel mode share estimation. In future research, enriching the training dataset that could cover different regions, such as the GPS-enhanced travel survey dataset available in Transportation Secure Data Center (TSDC) at National Renewable Energy Lab (NREL), might help improve the performance of our proposed framework by considering the travel behavior heterogeneity. In addition, the proposed framework relies on multimodal transportation networks, including drive, rail and bus. For regions without well-maintained transportation networks, it could be hard to capture the rail/bus travel. To decrease the dependency on multimodal transportation networks, additional information such as acceleration and stop time can be potentially considered.

- **Validation Data:** The proposed framework provides a general way to estimate travel mode shares at different geographies, with drive mode well captured, bus and rail mode slightly underestimated and non-motorized mode slightly overestimated. To further improve the performance of the proposed framework and capture finer level multimodal travel trends, additional data (i.e., transit ridership data, station-based metro passenger volume data) could be collected as external validation source and control totals. In addition, the current heuristic rule filters air trips according to domain knowledge, such as thresholds of average travel speed, trip time and trip distance, which could be refined with a ground-truth data collection for long-distance travels. A similar data collection effort to the procedure done by this paper could address the limitation and is on our research agenda.

- **Sample Bias:** In the two case studies, the travel mode share is estimated from the LBS data with a sample of the population, which might not be able to represent the population-level travel behavior. Also, the LBS data could not capture the travel behaviors of the population without mobile devices, which might yield an underrepresentation of the younger, elder, and low-income population. To address these two problems, an additional weighting and validation process can be done on top of the sample results using land use and sociodemographic information.

Our proposed framework is able to provide a timely and continuous measurement of travel trends and travel mode shares as a supplement to travel surveys. To achieve better estimates in practice, our proposed framework should be firstly calibrated using the travel surveys and then be applied to the MDLD datasets for preliminary investigation of the latest travel trends and mode shares. Although the proposed framework is not ready to completely replace travel surveys, it could help the government and local agencies refine the design of their travel surveys after prioritizing their data needs and before investing time and money in actual implementation. In addition, the proposed framework can also be applied to other realms, such as business development and public health. For instance, during the COVID-19 pandemic, a handful of research utilized the MDLDs to derive the travel statistics, i.e., trip rate and inflow, and study their correlation with the new COVID-19 infections (Xiong et al. 2020; Xiong et al. 2020; Hu et al. 2020; Xiao et al. 2020; Pan et al. 2020). By applying our proposed framework to the MDLDs, the correlation between multimodal travel and the spread of COVID-19 can be studied to support governments decision makings on transit or airline operations.

**Declarations**

**Funding**

The research is partially financially supported by a USDOT Federal Highway Administration (FHWA) Exploratory Advanced Research (EAR) project entitled "Data Analytics and Modeling Methods for Tracking and Predicting Origin-Destination Travel Trends based on Mobile Device Data" (Award No. 693JJ31750013). The opinions in this paper do not necessarily reflect the official views of USDOT or FHWA.

**Conflict of interest**

The authors declare that they have no conflict of interest.

**Availability of data and material**

Not applicable.

**Code availability**

Not applicable.

**Authors' contributions:**

M.Y., Y.P., and L.Z. designed research; M.Y., A.D. and S.G. processed data; M.Y., Y.P., and C.X. developed the algorithms and models; M.Y., A.D. and S.G. conducted case studies; M.Y., Y.P., A.D., S.G., C.X., L.Z. wrote the paper.

## Appendix I. Travel Mode Imputation with Machine Learning Models

### I.1. Machine Learning Overview.

KNN is one of the earliest and simplest classification models (Peterson 2009). The main idea of KNN is to find the top $k$ nearest samples of the target sample based on distance measurement. The distance between two samples is usually calculated based on Euclidean distance. SVC was developed by Cortes and Vapnik in the 1990s and applied for face recognition, pattern recognition, etc (Cortes and Vapnik 1995; Osuna et al. 1997; Suykens and Vandewalle 1999; Wang 2005). SVC can address the non-linearly separable samples by using the kernel function to map the data into a higher dimension, thus finding a hyperplane that best divides the data into different classes. Some examples of the kernel functions include polynomial, Gaussian, Gaussian radial basis function (RBF), sigmoid, etc. XGB is one of the most recent ensemble-learning algorithms using the boosting technique (Chen et al. 2015; Chen and Carlos 2016). The main idea of boosting is to train a set of weak classifiers using the same samples and then combine them into one strong classifier to improve the classification accuracy, where new classifiers are added to reduce errors based on previous models until no further improvements can be made (Kearns and Valiant 1988; Michael and Valiant 1994). RF is one of the most famous ensemble-learning algorithms using the bagging technique (Liaw and Wiener 2002; Breiman 1996). Bagging (Bootstrap aggregating) is a machine learning technique that tends to improve the stability and accuracy by generating multiple training sets by sampling from the data uniformly and with replacement (Breiman 1996). RF not only employs the bagging technique but also used a modified tree learning algorithm that selects a random subset of the features without using all features, which is called feature bagging (Liaw and Wiener 2002). In short, RF is essentially a collection of decision trees (Quinlan 1986) and each decision tree is trained with the different training sets and different features. The classification result follows the majority vote of all the decision trees in the forest. DNN is an Artificial Neural Network (ANN) with multiple layers between the input and output layers (Bengio 2009; Schmidhuber 2015). DNN can model the complex non-linear relationship between the input and the output by updating the weight vertices connecting each virtual neural between layers through back-propagation (Hecht-Nielsen 1992). Detailed methodology of DNN and ANN can be found in Bengio (2009) and Schmidhuber (2015).

## I.2. Fine-Tuned Parameters for Each Machine Learning Method

Table 6. Fine-Tuned Hyperparameters for Different Machine Learning Models.

| Four Travel Modes | | Five Travel Modes | |
|---|---|---|---|
| Model | Parameters | Model | Parameters |
| KNN | k = 5;<br>p = 1;<br>weights = 'distance';<br>algorithm = 'brute';<br>leaf_size = 1;<br>metric = 'minkoski'. | KNN | k = 1;<br>p = 1;<br>weights = 'distance';<br>algorithm = 'ball_tree';<br>leaf_size = 1;<br>metric = 'minkoski'. |
| SVC | Cs = 100;<br>Gammas = 1;<br>class_weight = 'None';<br>kernel = 'rbf'. | SVC | Cs = 100;<br>Gammas = 1;<br>class_weight = 'None';<br>kernel = 'rbf'. |
| RF | n_estimators = 700;<br>max_features = 'sqrt';<br>max_depth = 30;<br>min_samples_split = 2;<br>min_samples_leaf = 1;<br>bootstrap = 'False';<br>class_weight = 'None'. | RF | n_estimators = 800;<br>max_features = 'auto';<br>max_depth = 60;<br>min_samples_split = 2;<br>min_samples_leaf = 1;<br>bootstrap = 'False';<br>class_weight = 'balanced_subsample'. |
| XGB | colsample_bytree = 0.980;<br>gamma = 0.429;<br>learning_rate = 0.317;<br>max_depth = 5;<br>n_estimators = 136;<br>subsample = 0.627. | XGB | colsample_bytree = 0.865;<br>gamma = 0.299;<br>learning_rate = 0.324;<br>max_depth = 5;<br>n_estimators = 126;<br>subsample = 0.884. |
| DNN | learning_rate = 0.001;<br>epochs = 500;<br>batch_size = 32;<br>optimizer: rmsprop;<br>loss; 'categorical_crossentropy';<br>layers: 64-32-16-8-DropOut (0.2). | DNN | learning_rate = 0.001;<br>epochs = 500;<br>batch_size = 32;<br>optimizer: rmsprop;<br>loss; 'categorical_crossentropy';<br>layers: 64-32-16-8-DropOut (0.2). |

### I.3. Feature Importance of the Random Forest Model

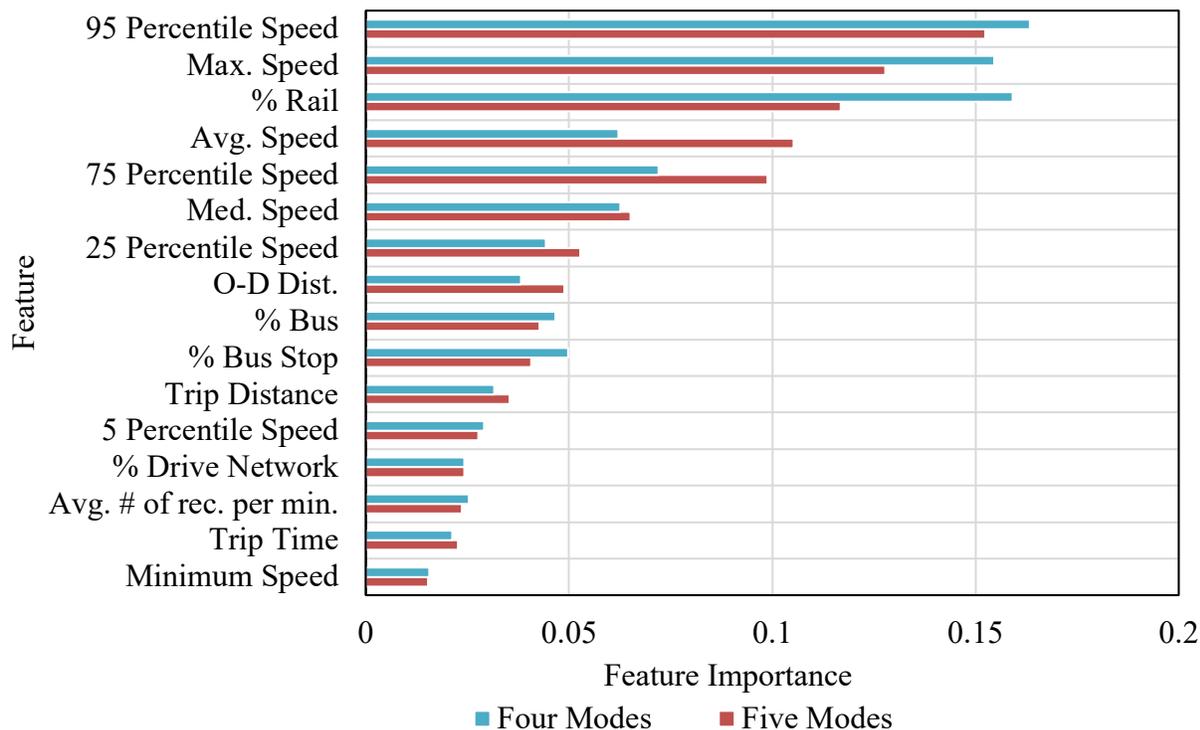

Figure 15. Feature Importance of the Random Forest Classifier.

Figure 15 shows the feature importance value of the RF model for four and five travel modes respectively. The feature importance value (Gini importance) is automatically calculated using the python package scikit-learn (Pedregosa et al. 2011), representing each importance as the sum over the number of splits across all trees. It can be seen that the speed variables (95 quantile speed, maximum speed, and average speed) are the most important. In addition, the percentage of records that fell within 50 meters of the rail network is also significantly important since it is a representative feature to impute rail trips.

# Appendix II. Spatial Accuracy and Location Recording Interval of the Location-Based Service Data

Table 7 shows the detailed spatial accuracy distribution of the three datasets. It can be clearly seen that the incenTrip dataset has the highest data quality among the three datasets, with over 80% of the data has a spatial accuracy smaller than 50-meter. For the other two large-scale LBS datasets, the spatial accuracy shows a similar bimodal distribution, with the second peak locates around 70-meter. In general, for all three datasets, around 90% of the data has spatial accuracy smaller than 100-meter. Therefore, before we apply our proposed framework to the dataset, we removed the location data with spatial accuracy larger than 100-meter for all three datasets to ensure the data quality.

Table 7. Spatial Accuracy Distribution of the Three Datasets

| Spatial Accuracy | incenTrip | | Dataset I | | Dataset II | |
|---|---|---|---|---|---|---|
| | Proportion | Cumulation | Proportion | Cumulation | Proportion | Cumulation |
| (0, 10] | 44.15% | 44.15% | 49.91% | 49.91% | 50.21% | 50.21% |
| (10, 20] | 22.50% | 66.64% | 7.58% | 57.49% | 5.78% | 55.99% |
| (20, 30] | 8.79% | 75.43% | 6.80% | 64.28% | 7.27% | 63.27% |
| (30, 40] | 4.29% | 79.73% | 1.00% | 65.28% | 1.07% | 64.33% |
| (40, 50] | 2.17% | 81.90% | 3.47% | 68.76% | 4.10% | 68.43% |
| (50, 60] | 2.01% | 83.92% | 0.26% | 69.01% | 0.28% | 68.71% |
| (60, 70] | 1.10% | 85.01% | 18.33% | 87.34% | 17.60% | 86.31% |
| (70, 80] | 0.92% | 85.93% | 0.69% | 88.04% | 0.83% | 87.14% |
| (80, 90] | 0.79% | 86.73% | 0.66% | 88.70% | 0.80% | 87.94% |
| (90, 100] | 1.19% | 87.92% | 0.59% | 89.29% | 0.66% | 88.60% |
| (100, 200] | 3.81% | 91.73% | 4.18% | 93.46% | 4.87% | 93.47% |
| (200, 500] | 2.57% | 94.31% | 0.30% | 93.77% | 0.23% | 93.70% |
| (500, | 5.69% | 100.00% | 6.23% | 100.00% | 6.30% | 100.00% |

Table 8 shows the detailed LRI distributions for the two large-scale LBS datasets after removing the location data with spatial accuracy larger than 100-meter, which also follows the bimodal distribution, with the second peak locates around 120-s. In both datasets, more than 50% of the data has an LRI smaller than 30 seconds.

Table 8. Location Recording Frequency of the Two Case Studies' Datasets

| LRI | Dataset I | | Dataset II | |
|---|---|---|---|---|
| | Proportion | Cumulation | Proportion | Cumulation |
| (0, 10] | 44.81% | 44.81% | 46.48% | 46.48% |
| (10, 20] | 4.98% | 49.79% | 4.92% | 51.40% |
| (20, 30] | 1.83% | 51.62% | 1.77% | 53.17% |
| (30, 40] | 1.21% | 52.83% | 1.15% | 54.32% |
| (40, 50] | 0.86% | 53.69% | 0.81% | 55.13% |
| (50, 60] | 0.73% | 54.42% | 0.68% | 55.81% |
| (60, 70] | 0.64% | 55.06% | 0.60% | 56.42% |
| (70, 80] | 0.50% | 55.56% | 0.48% | 56.90% |
| (80, 90] | 0.45% | 56.01% | 0.43% | 57.32% |
| (90, 100] | 0.38% | 56.39% | 0.36% | 57.69% |
| (100, 150] | 6.46% | 62.85% | 6.70% | 64.39% |
| (150, 200] | 2.78% | 65.63% | 2.75% | 67.14% |
| (200, 250] | 1.11% | 66.73% | 1.08% | 68.22% |
| (250, 300] | 2.11% | 68.84% | 2.13% | 70.34% |
| (300, 350] | 1.33% | 70.17% | 1.30% | 71.64% |
| (350, 400] | 0.75% | 70.92% | 0.75% | 72.40% |
| (400, 500] | 1.06% | 71.98% | 1.06% | 73.46% |
| (500, | 28.02% | 100.00% | 26.54% | 100.00% |

**Appendix III. Travel Mode Share Estimation Results**

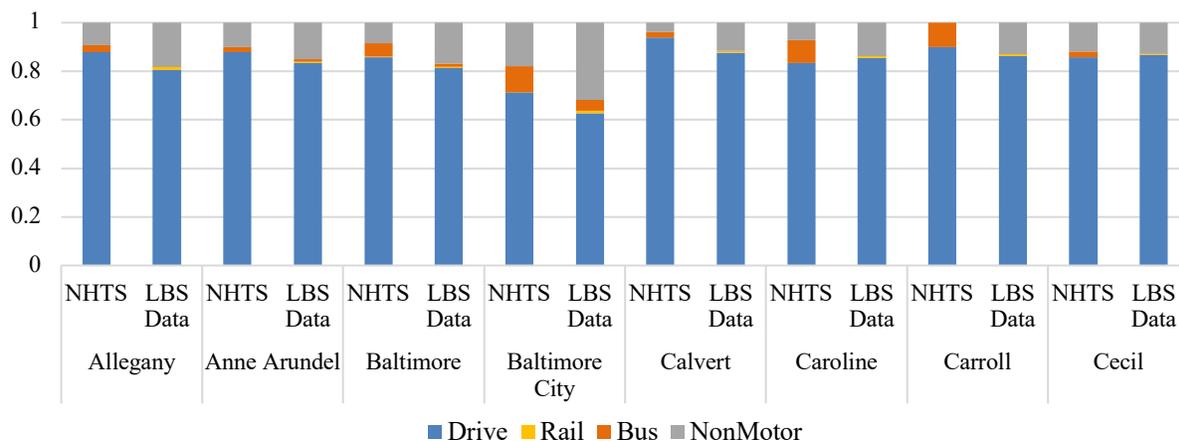

(a)

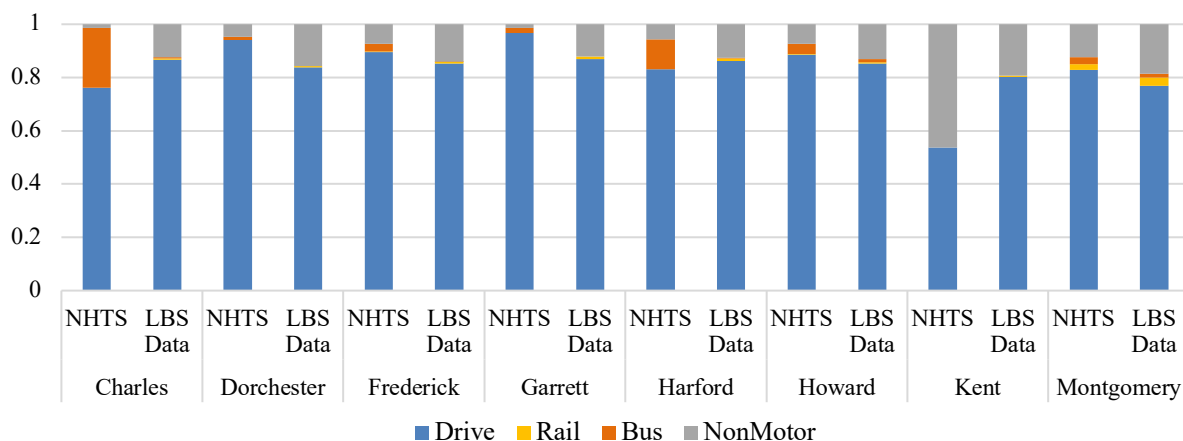

(b)

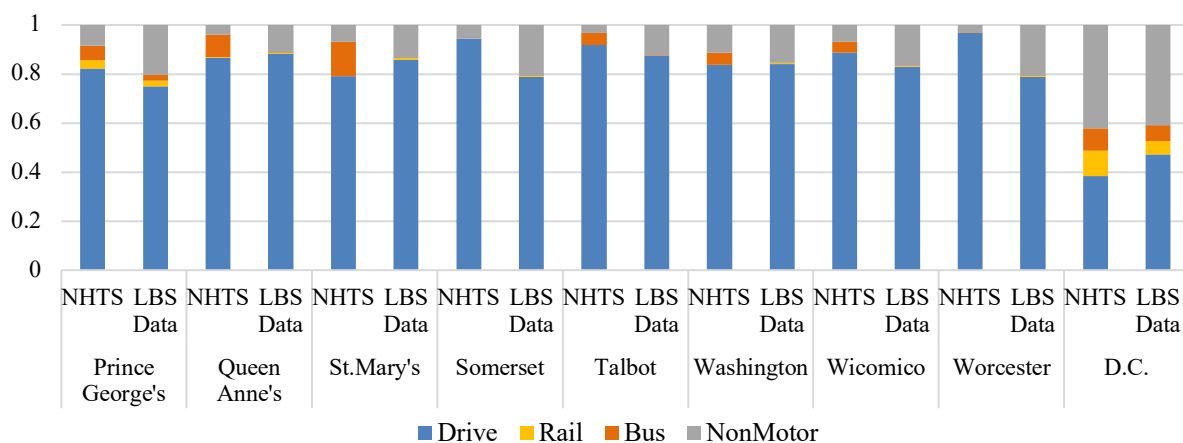

(c)

Figure 16. County-level travel mode share comparison between Dataset I and NHTS 2017.

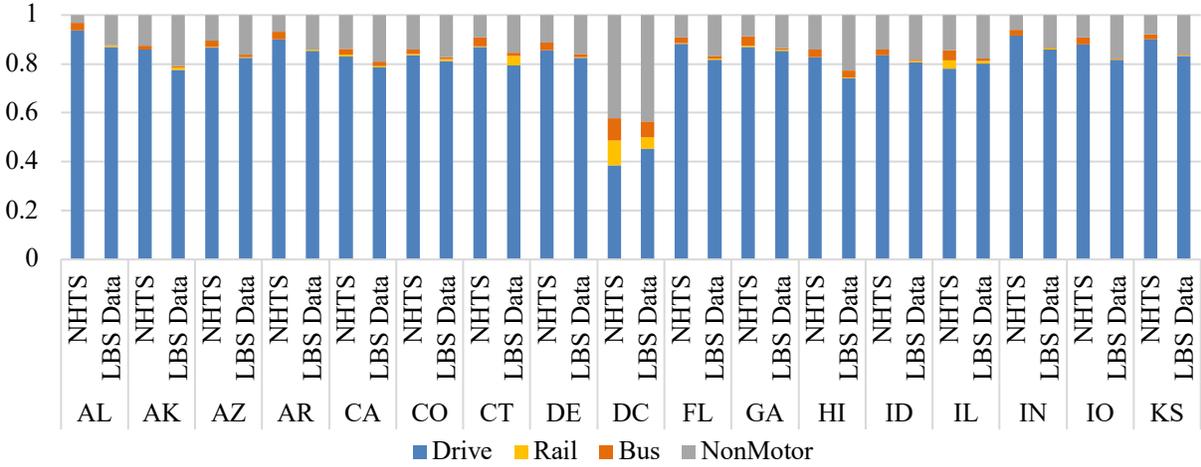

(a)

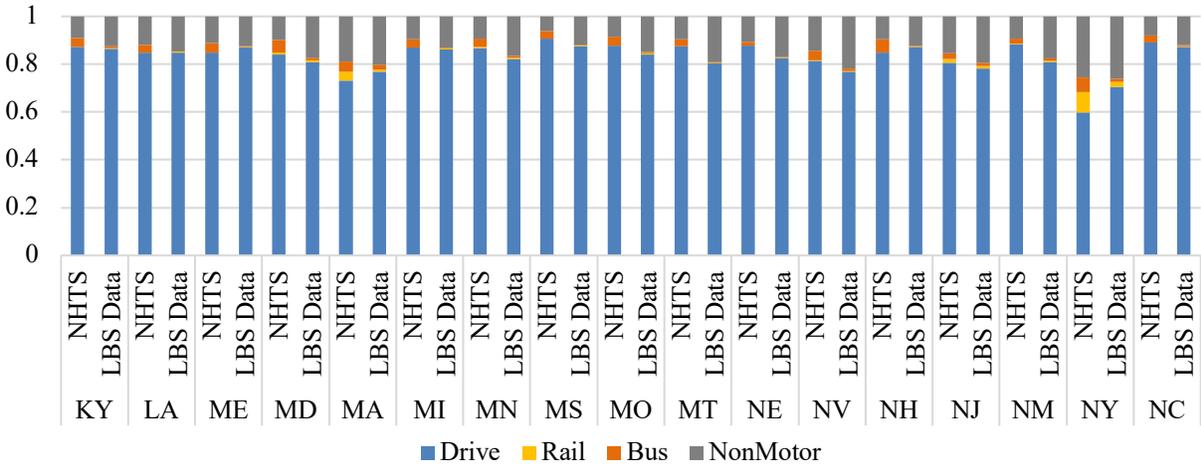

(b)

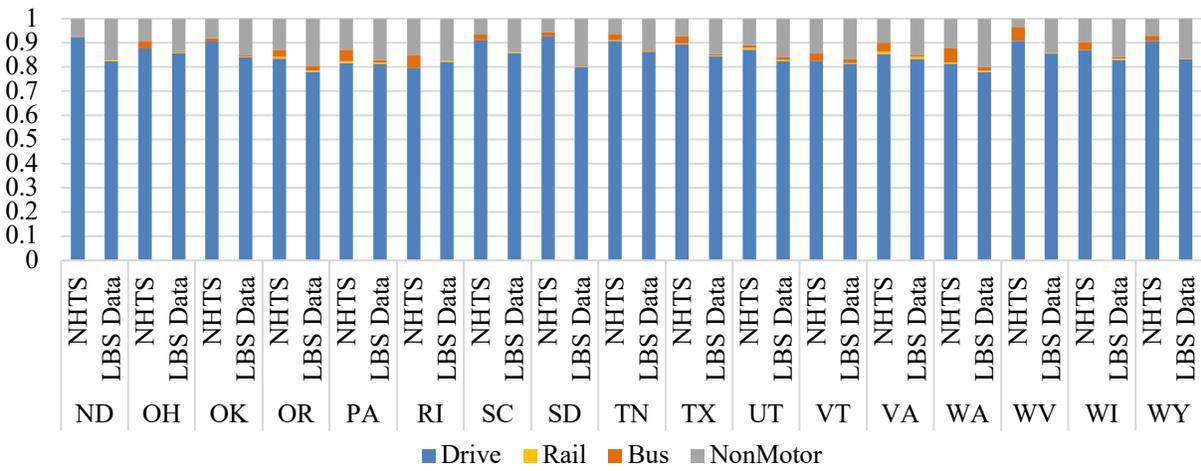

(c)

Figure 17. State-level travel mode share comparison between Dataset II and NHTS 2017.

Table 9. County-level travel mode share comparison between Dataset I and 2007/08 TPB-BMC HHTS.

| County | 2007-08 TPB-BMC HHTS | | | | LBS Data | | | |
|---|---|---|---|---|---|---|---|---|
| | Drive | Rail | Bus | NonMotor | Drive | Rail | Bus | NonMotor |
| Anne Arundel | 92.285% | 0.728% | 3.666% | 3.321% | 83.351% | 0.694% | 1.030% | 14.925% |
| Baltimore | 90.773% | 0.741% | 4.106% | 4.379% | 81.234% | 0.628% | 1.324% | 16.814% |
| Baltimore City | 71.962% | 2.555% | 9.163% | 16.320% | 62.658% | 1.140% | 4.383% | 31.819% |
| Carroll | 92.411% | 0.361% | 4.705% | 2.523% | 86.340% | 0.704% | 0.050% | 12.906% |
| DC | 47.160% | 14.225% | 7.533% | 31.083% | 47.129% | 5.532% | 6.568% | 40.771% |
| Harford | 91.646% | 0.115% | 3.891% | 4.348% | 86.221% | 0.876% | 0.253% | 12.650% |
| Howard | 90.461% | 0.464% | 5.294% | 3.780% | 85.274% | 0.441% | 1.096% | 13.189% |
| Montgomery | 82.504% | 2.719% | 5.749% | 9.028% | 76.979% | 2.917% | 1.446% | 18.659% |
| Prince George's | 85.244% | 2.856% | 6.474% | 5.425% | 75.069% | 2.187% | 2.255% | 20.489% |

Table 10. County-level travel mode share comparison between Dataset I with NHTS 2017.

| County | 2007-08 TPB-BMC HHTS | | | | LBS Data | | | |
|---|---|---|---|---|---|---|---|---|
| | Drive | Rail | Bus | NonMotor | Drive | Rail | Bus | NonMotor |
| Allegany | 87.807% | 0.000% | 2.959% | 9.234% | 80.437% | 1.405% | 0.104% | 18.054% |
| Anne Arundel | 87.724% | 0.016% | 2.235% | 10.024% | 83.351% | 0.694% | 1.030% | 14.925% |
| Baltimore | 85.655% | 0.403% | 5.386% | 8.556% | 81.234% | 0.628% | 1.324% | 16.814% |
| Baltimore City | 71.348% | 0.095% | 10.556% | 18.001% | 62.658% | 1.140% | 4.383% | 31.819% |
| Calvert | 93.816% | 0.000% | 2.456% | 3.728% | 87.470% | 0.747% | 0.222% | 11.562% |
| Caroline | 83.464% | 0.000% | 9.353% | 7.183% | 85.516% | 0.781% | 0.087% | 13.617% |
| Carroll | 89.947% | 0.000% | 10.053% | 0.000% | 86.340% | 0.704% | 0.050% | 12.906% |
| Cecil | 85.739% | 0.000% | 2.311% | 11.950% | 86.746% | 0.636% | 0.021% | 12.598% |
| Charles | 76.195% | 0.000% | 22.494% | 1.311% | 86.656% | 0.510% | 0.382% | 12.452% |
| DC | 94.103% | 0.000% | 1.249% | 4.648% | 47.129% | 5.532% | 6.568% | 40.771% |
| Dorchester | 89.483% | 0.218% | 3.028% | 7.271% | 83.926% | 0.464% | 0.000% | 15.611% |
| Frederick | 96.711% | 0.000% | 1.845% | 1.444% | 85.190% | 0.750% | 0.123% | 13.938% |
| Garrett | 83.215% | 0.000% | 11.136% | 5.649% | 87.016% | 0.749% | 0.000% | 12.235% |
| Harford | 88.492% | 0.296% | 3.917% | 7.295% | 86.221% | 0.876% | 0.254% | 12.650% |
| Howard | 53.718% | 0.000% | 0.000% | 46.282% | 85.274% | 0.441% | 1.096% | 13.189% |
| Kent | 82.855% | 2.218% | 2.666% | 12.261% | 80.349% | 0.465% | 0.000% | 19.186% |
| Montgomery | 82.108% | 3.363% | 6.060% | 8.469% | 76.979% | 2.917% | 1.446% | 18.659% |
| Prince George's | 86.707% | 0.088% | 9.284% | 3.921% | 75.069% | 2.187% | 2.255% | 20.489% |
| Queen Anne's | 79.223% | 0.000% | 13.979% | 6.798% | 88.069% | 0.558% | 0.147% | 11.225% |
| Somerset | 94.433% | 0.000% | 0.000% | 5.567% | 78.926% | 0.403% | 0.134% | 20.537% |
| St. Mary's | 91.749% | 0.070% | 4.866% | 3.315% | 85.908% | 0.718% | 0.112% | 13.262% |
| Talbot | 83.786% | 0.098% | 4.813% | 11.303% | 86.770% | 0.495% | 0.106% | 12.628% |
| Washington | 88.606% | 0.000% | 4.395% | 7.000% | 84.030% | 0.506% | 0.000% | 15.465% |
| Wicomico | 96.828% | 0.000% | 0.209% | 2.962% | 82.988% | 0.348% | 0.070% | 16.594% |
| Worcester | 38.446% | 10.277% | 9.038% | 42.239% | 78.947% | 0.301% | 0.030% | 20.722% |

Table 11. State-level travel mode share comparison between Dataset II and NHTS 2017.

| State | NHTS 2017 | | | | LBS Data | | | |
|---|---|---|---|---|---|---|---|---|
| | Drive | Rail | Bus | NonMotor | Drive | Rail | Bus | NonMotor |
| AL | 93.730% | 0.000% | 3.228% | 3.042% | 86.991% | 0.365% | 0.257% | 12.387% |
| AK | 85.807% | 0.000% | 1.555% | 12.638% | 77.536% | 1.034% | 0.447% | 20.984% |
| AZ | 86.671% | 0.261% | 2.687% | 10.381% | 82.416% | 0.458% | 0.856% | 16.270% |
| AR | 90.110% | 0.000% | 3.152% | 6.737% | 85.397% | 0.367% | 0.246% | 13.990% |
| CA | 82.996% | 0.877% | 2.391% | 13.736% | 78.516% | 0.696% | 1.765% | 19.024% |
| CO | 83.664% | 0.291% | 2.051% | 13.994% | 81.318% | 0.637% | 0.778% | 17.267% |
| CT | 87.004% | 0.270% | 3.607% | 9.119% | 79.353% | 3.968% | 1.229% | 15.449% |
| DE | 85.623% | 0.000% | 3.052% | 11.325% | 82.412% | 0.421% | 1.244% | 15.923% |
| DC | 38.446% | 10.277% | 9.038% | 42.239% | 45.271% | 4.620% | 6.551% | 43.558% |
| FL | 88.334% | 0.139% | 2.252% | 9.276% | 81.840% | 0.352% | 1.035% | 16.772% |
| GA | 87.039% | 0.432% | 3.891% | 8.637% | 85.424% | 0.421% | 0.536% | 13.619% |
| HI | 82.758% | 0.000% | 3.081% | 14.162% | 74.049% | 0.318% | 3.007% | 22.625% |
| ID | 83.426% | 0.000% | 2.374% | 14.199% | 80.734% | 0.418% | 0.533% | 18.315% |
| IL | 78.180% | 3.295% | 4.064% | 14.460% | 80.112% | 1.216% | 1.021% | 17.650% |
| IN | 91.515% | 0.027% | 2.421% | 6.037% | 85.844% | 0.472% | 0.302% | 13.382% |
| IO | 88.007% | 0.000% | 2.712% | 9.280% | 81.602% | 0.415% | 0.202% | 17.782% |
| KS | 90.171% | 0.000% | 1.872% | 7.957% | 83.279% | 0.400% | 0.250% | 16.070% |
| KY | 87.221% | 0.000% | 3.837% | 8.942% | 86.451% | 0.293% | 0.659% | 12.597% |
| LA | 84.754% | 0.009% | 3.173% | 12.064% | 84.861% | 0.437% | 0.042% | 14.660% |
| ME | 84.901% | 0.000% | 3.924% | 11.175% | 86.913% | 0.352% | 0.171% | 12.563% |
| MD | 83.926% | 0.904% | 5.396% | 9.774% | 80.768% | 0.868% | 1.061% | 17.303% |
| MA | 73.094% | 3.804% | 4.156% | 18.945% | 76.801% | 0.932% | 2.183% | 20.084% |
| MI | 86.918% | 0.066% | 3.434% | 9.582% | 86.202% | 0.410% | 0.245% | 13.143% |
| MN | 86.800% | 0.528% | 3.461% | 9.211% | 82.265% | 0.460% | 0.704% | 16.571% |
| MS | 90.762% | 0.000% | 3.260% | 5.978% | 87.509% | 0.415% | 0.060% | 12.016% |
| MO | 87.743% | 0.048% | 3.744% | 8.465% | 84.161% | 0.408% | 0.562% | 14.868% |
| MT | 87.424% | 0.000% | 3.060% | 9.516% | 80.275% | 0.424% | 0.180% | 19.122% |
| NE | 87.648% | 0.000% | 1.800% | 10.551% | 82.419% | 0.387% | 0.144% | 17.050% |
| NV | 81.382% | 0.306% | 4.043% | 14.269% | 76.935% | 0.230% | 0.954% | 21.881% |
| NH | 84.751% | 0.000% | 5.688% | 9.561% | 87.187% | 0.333% | 0.299% | 12.181% |
| NJ | 80.564% | 1.541% | 2.591% | 15.303% | 78.124% | 1.221% | 1.324% | 19.330% |
| NM | 88.419% | 0.109% | 2.199% | 9.273% | 80.942% | 0.518% | 1.008% | 17.531% |
| NY | 59.810% | 8.620% | 5.983% | 25.586% | 70.472% | 2.194% | 1.342% | 25.992% |
| NC | 89.009% | 0.063% | 3.037% | 7.891% | 87.275% | 0.361% | 0.281% | 12.084% |
| ND | 92.371% | 0.000% | 0.419% | 7.210% | 82.452% | 0.465% | 0.050% | 17.033% |
| OH | 87.748% | 0.018% | 3.163% | 9.071% | 85.945% | 0.296% | 0.439% | 13.320% |
| OK | 90.560% | 0.000% | 1.462% | 7.978% | 84.098% | 0.294% | 0.284% | 15.324% |
| OR | 83.564% | 0.604% | 2.745% | 13.087% | 77.947% | 0.538% | 1.964% | 19.551% |
| PA | 81.608% | 0.831% | 4.727% | 12.833% | 81.196% | 0.628% | 1.176% | 16.999% |
| RI | 79.818% | 0.024% | 4.859% | 15.299% | 82.179% | 0.399% | 0.080% | 17.342% |
| SC | 91.041% | 0.026% | 2.392% | 6.542% | 85.958% | 0.307% | 0.224% | 13.511% |
| SD | 92.903% | 0.102% | 1.488% | 5.507% | 79.965% | 0.371% | 0.283% | 19.380% |
| TN | 90.954% | 0.077% | 2.716% | 6.254% | 86.290% | 0.322% | 0.438% | 12.950% |
| TX | 89.527% | 0.205% | 3.018% | 7.250% | 84.423% | 0.371% | 0.529% | 14.678% |
| UT | 87.304% | 0.529% | 1.275% | 10.892% | 82.325% | 0.528% | 1.039% | 16.108% |
| VT | 82.510% | 0.000% | 3.077% | 14.413% | 81.295% | 0.344% | 1.575% | 16.786% |
| VA | 85.252% | 1.169% | 3.716% | 9.863% | 83.307% | 0.887% | 0.714% | 15.093% |
| WA | 81.362% | 0.580% | 5.645% | 12.413% | 77.979% | 0.511% | 1.490% | 20.020% |
| WV | 90.802% | 0.000% | 5.759% | 3.439% | 85.569% | 0.473% | 0.139% | 13.819% |
| WI | 87.135% | 0.013% | 3.172% | 9.679% | 82.889% | 0.551% | 0.674% | 15.886% |
| WY | 90.873% | 0.000% | 2.204% | 6.923% | 83.101% | 0.263% | 0.020% | 16.616% |